\documentclass[prm,notitlepage,twocolumn]{revtex4-2}
\usepackage{chemformula} 
\usepackage[T1]{fontenc} 
\usepackage{amsmath}
\usepackage{amssymb}
\usepackage{graphicx}
\usepackage{subcaption}
\usepackage{booktabs}
\usepackage{multirow}
\usepackage{tablefootnote}
\usepackage{float}
\usepackage{relsize}
\usepackage[version=3]{mhchem}
\usepackage{dcolumn}
\newcolumntype{.}{ D{.}{.}{-1} }
\usepackage{setspace}
\usepackage{ulem}
\usepackage{booktabs}
\usepackage{xcolor}
\usepackage{hyperref}
\usepackage{underscore}
\usepackage{comment}

\newcommand{\beq}{\begin{equation}}
\newcommand{\eeq}{\end{equation}}
\newcommand{\bea}{\begin{eqnarray}}
\newcommand{\eea}{\end{eqnarray}}

\def\br{{\mathbf{r}}}

\newcommand{\benum}{\begin{enumerate}}
\newcommand{\eenum}{\end{enumerate}}
\newcommand{\bi}{\begin{itemize}}
\newcommand{\ei}{\end{itemize}}
\def\bfr{{\mathbf{r}}}

\def\sss{\scriptscriptstyle\rm}

\def\x{_{\sss X}}
\def\c{_{\sss C}}
\def\s{_{\sss S}}
\def\xc{_{\sss XC}}

\def\TF{_{\sss TF}}

\def\vW{^{\sss W}}
\def\svW{_{\sss W}}
\def\SV{^{\sss SV}}
\def\OF{_{\sss OF}}
\def\SOF{^{\sss OF}}
\def\PC{_{\sss PC}}
\def\RPP{_{\sss RPP}}

\def\CR{_{\sss CR}}
\def\SRPP{_{\sss SRPP}}
\def\LDA{^{\sss LDA}}
\def\rtwoSCAN{^{\sss r^2 SCAN}}

\def\opt{$_\mathrm{opt}$}

\DeclareUnicodeCharacter{2212}{-}

\begin{document}

\author{H. Francisco}
\affiliation{Quantum Theory Project, Dept.\ of Physics, University of Florida, Gainesville FL 32611, USA}
\email{francisco.hector@ufl.edu}

\author{B. Thapa}
\affiliation{Dept.\ of Physics and Astronomy, George Mason University, Fairfax VA 22030, USA}
\email{bthapa3@gmu.edu}

\author{S.B. Trickey}
\affiliation{Quantum Theory Project, Dept.\ of Physics and Dept.\ of Chemistry,  University of Florida, Gainesville FL 32611, USA}
\email{trickey@ufl.edu}

\author{A.C. Cancio}
\affiliation{Dept.\ of Physics and Astronomy, Ball State University, Muncie IN 47306, USA}

\title{Performance Improvement of Deorbitalized Exchange-Correlation Functionals}

\date{21 Jan. 2026; 07 Feb. 2026}


\begin{abstract}
Deorbitalization of a conventional meta-generalized-gradient
exchange-correlation approximation replaces its dependence
upon the Kohn-Sham kinetic energy density with a dependence on the
density gradient and Laplacian. In principle, that simplification
should provide improved computational performance relative to the
original meta-GGA form because of the shift from an orbital-dependent
generalized Kohn-Sham potential to a true KS local potential. Often
that prospective gain is lost because of problematic roughness in the
density caused by the density Laplacian and consequent roughness in
the exchange-correlation potential from the resulting higher-order
spatial derivatives of the density in it.  We address the problem
by constructing a deorbitalizer based on the RPP deorbitalizer
[Phys.\ Rev.\ Mater.\ \textbf{6}, 083803 (2022)] with comparative
smoothness of the potential along with retention of constraint satisfaction
as design goals.  Applied to the
r$^2$SCAN exchange-correlation functional
[J.\ Phys.\ Chem.\ Lett.\ \textbf{11}, 8208 (2020)], we find substantial
timing improvements for solid-state calculations over both r$^2$SCAN
and its earlier deorbitalization for high precision calculations of
structural properties, while improving upon the accuracy of RPP
deorbitalization for both solids and molecules.
\end{abstract}

\maketitle

\section{Introduction \label{sec:intro}}

The enormous impact of Hohenberg-Kohn-Sham density functional theory
(DFT) upon computational study of molecular and materials properties
is a consequence of the remarkable cost-accuracy balance of DFT when
implemented with modern approximate exchange-correlation (XC)
functionals.  The cost-accuracy trade-off is determined by the
intrinsic limits and transferability of the specific XC approximation
used.  That compromise clearly differs by problem class.  A small set
of very large molecules can be treated with an XC approximation far up
the Perdew-Schmidt ``Jacob's ladder'' complexity hierarchy
\cite{PerdewSchmidt}. In contrast, the computational costs of
high-throughput screening calculations on large data sets of such
molecules or ab initio molecular dynamics (AIMD) on counterpart
condensed phases compel use of lower-rung approximations.  See, for
example, the discussion in Sections 2.2.21-23
of Ref. \citenum{DFTroundtable}.

The SCAN meta generalized-gradient approximation (meta-GGA)
\cite{SCAN,SCANNature} and its variants \cite{FurnessSun19,BartokYates19,FurnessEtAl2020}, 
have proven to be effective compromises for accuracy and efficiency
in structural and energetic 
calculations for both solids and molecules ~\cite{Zhang2018,SCANNature} albeit with
some limitations \cite{KTBM19,SCANOverMag}. 
A design advantage of such meta-GGA functionals is the introduction of 
inherently nonlocal density information through the use of the non-interacting 
(Kohn-Sham, KS) kinetic energy density~\cite{BeckeEdgecombe,SunEtAl2013},
\beq
\tau\s  :=  \frac{1}{2}\sum_i f_i |\nabla \phi_i(\bfr)|^2
\label{eq:tausdefn}
\eeq
with orbitals $\phi_i$ and occupation numbers $f_i$.  [Throughout we 
consider only non-relativistic and non-spin-polarized ($f_{i} = 0,1,2$) systems and use Hartree atomic units unless otherwise noted.]
A resultant design
advantage is satisfaction of many more exact constraints than possible
with a GGA, particularly for the iso-orbital limit.

Despite the successes, there are difficulties of implementation with
these meta-GGAs that limit their full exploitation.  The explicit
orbital-dependence introduced by $\tau\s$ causes typical practical
application to rely on non-local XC potentials in generalized KS (gKS)
equations. Solutions to those can be slow relative to pure KS
implementations~\cite{SCANL1}.  This slowness also originates with the
complicated form of the functionals and their potentials, their
consequent sensitivity to the density~\cite{PAWS22}, and concomitant
need for dense grids for converged
integrals~\cite{YaoKanai2017,Lehtola22}. Calculations sometimes are
done non-self-consistently,~\cite{Zhang2018, PartDeorb24}. 
Projector-augmented wave (PAW) effective potentials for these functionals have proven difficult to
develop \cite{PAWS22}.

Effort to resolve these issues has led to the redesign of SCAN, first
by streamlining performance at the cost of significant constraint
noncompliance,~\cite{BartokYates19} then by iso-orbital indicator
redefinition,~\cite{FurnessSun19} and, most recently, in
r$^2$SCAN,~\cite{FurnessEtAl2020} by restoration of most of the
constraints met by SCAN, while retaining the improvements in numerical
performance of prior variants.  Nonetheless, some numerical problems
remain \cite{PAWS22, Lehtola22}.


Deorbitalization~\cite{SCANL1,SCANL2,r2SCANL} 
has been developed as a strategy both to obviate numerical issues 
stemming from use of gKS and to retain 
the interpretive power of the pure KS equation. In deorbitalization, 
the $\tau\s$ dependence used in meta-GGA XC functionals  as a local indicator of  bond character is replaced  
with a function of the  density $n(\br)$, its gradient $\nabla n$, and  
Laplacian $\nabla^2 n$.  The result is a pure XC density functional and a pure
(local) KS potential.  

The obvious deorbitalization requirement is a sufficiently accurate approximate
orbital-free kinetic energy density (KED) functional. That is an
active area of research.~\cite{OFKEreview} 
A usefully accurate description of the KED for deorbitalizing a
meta-GGA indicator functional by a simple generalized gradient
approximation (GGA) alone does not seem possible \cite{SCANL1}.
Inclusion of the density Laplacian proves essential for accurate orbital-free 
description of the KED at a local level, as demonstrated with the  
KED density functional provided by Perdew and 
Constantin~\cite{PerdewConstantin2007} (and also Ref.~\citenum{CSK16}).
Practical deorbitalization
has come through the methodology of Mej\'ia-Rodr\'iguez and
Trickey~\cite{SCANL1,SCANL2,r2SCANL} (M-RT).  They reparametrized
existing Laplacian-level KED functionals, e.g. Perdew-Constantin
(PC)~\cite{PerdewConstantin2007} and others from orbital-free
DFT~\cite{OFKEreview}, to match KED-dependent indicator functions of
the first 18 isolated atoms.  The most generally successful case of
M-RT deorbitalization involves the reparametrization (PC\opt) of
original PC.  That gave successful deorbitalizations of both
SCAN~\cite{SCANL1, SCANL2} and r$^2$SCAN~\cite{r2SCANL}, in the sense
of providing what was deemed to be a faithful reproduction of the parent meta-GGA error
patterns for structural and energetic properties evaluated on standard
molecular and crystalline benchmarks. Such faithfulness in reproduction
defines a deorbitalization goal, one which we follow.

That faithfulness also has held up in application to materials and
outside the original test sets.~\cite{Rani_2024, Zhang_2024} Other
KEDs, e.g. Cancio-Redd (CR) ~\cite{CR} have led, for some meta-GGAs,~\cite{FCT25} to better error patterns than those produced with the exact KED, but that  has not been seen for best-in-breed meta-GGAs.

A limitation of the M-RT error-pattern criterion of deorbitalization
faithfulness is that it can allow acceptance of deorbitalized XC
functionals that deviate noticeably from the exact form in the limit
of slowly varying density, \textit{i.e.}, constraint violation.  In
the context of rethinking the PC functional form, the recent OFR2
study~\cite{OFR2} emphasized constraint satisfaction, including
restoration of correct slowly-varying limit behavior. That
recovers fourth-order gradient expansion compliance lost in going from
SCAN to r$^2$SCAN \cite{BartokYates19}. The resulting deorbitalization
does well in calculations for solids (retaining the numerical
advantages of rSCAN) but not as well as r$^2$SCAN-L (r$^2$SCAN
deorbitalized by PC\opt, the reparametrized version of PC used by
M-RT\cite{SCANL1,SCANL2}) for molecules.

As an aside for clarity, note that a second modification of the
original M-RT deorbitalization has been developed \cite{srTMpartIII}
to treat meta-GGAs that rely on more than one iso-orbital indicator,
particularly the Tao Mo functional (TM)~\cite{TM}. That is not
relevant here.

Alternative strategies to Laplacian-level deorbitalization include use
of higher-order spatial derivatives
\cite{deSilvaCorminboeuf2014,deSilvaCorminboeuf2015a,deSilvaCorminboeuf2015b}
and deorbitalization of the exchange potential
only~\cite{PartDeorb24}. Those are of only contextual interest here,
as they may have more significant numerical challenges than those
arising from the M-RT procedure.

Given success at reproducing the test-case performance of existing
orbital-dependent meta-GGAs, the challenge is the effective numerical
implementation of such a deorbitalized functional.  The well-known
sensitivity to density Laplacians serves as a caution.  See for but
one example, the discussion in Ref.\ \citenum{SCANL1}.
Laplacian-based XC deorbitalizers, like Laplacian-based KEDFs, can be
ill-behaved with regards to integral convergence and require
relatively dense grids.~\cite{Lehtola22}  Moreover, they also can
have large, spiky variations in the magnitude of the XC potential.
Such variations arise from the high-order spatial derivatives in the X potential
that follow from $\nabla^2 n$.~\cite{OFR2} 

Nonetheless, some of the M-RT functionals \cite{SCANL1,SCANL2} have
shown marked performance improvement over their gKS parents, with time
per self-consistent-field (SCF) cycle of the Kohn-Sham solution as
much as a factor of three shorter with the VASP code
\cite{kresse1996efficient,PhysRevB.59.1758}.  But recent work
indicates~\cite{srTMpartIII} that it is quite typical for
deorbitalized functionals to require many more self-consistent cycles
to achieve SCF convergence than for their parent functionals with gKS,
thus offsetting the gains made per cycle.

The natural, critical question raised by these issues is this: can
deorbitalization, in fact, succeed in achieving faster, more
reliable meta-GGA calculations?  At a more detailed level, how does the
difficulty of using $\nabla^2 n$ compare with the difficulties of
gKS calculations?  For which run conditions, and for which
applications, if any, does deorbitalization provide a clear advantage?
With respect to computational efficiency, where and how can M-RT
deorbitalization be improved?

One clear opportunity is to find ways to reduce numerical
instabilities induced by the density-Laplacian.  In this work, we take
a specific step in that amelioration by intervening in those aspects
of the deorbitalized functional that are most problematic, while
retaining (or even eventually restoring) important constraints along
the way. Our scheme is based on exploitation of the Cancio-Redd
\cite{CR} and Cancio-Stuart-Kuna \cite{CSK16} (CR hereafter)
approach of minimizing the structural complexity of KED functionals.
We combine that with the constraint satisfaction of the ``r$^2$SCAN
piecewise polynomial'' (hereafter RPP) deorbitalized indicator
function \cite{OFR2} used in the OFR2 functional to produce a
deorbitalizer with a smoother X potential than OFR2.  We show that
this functional performs well compared to r$^2$SCAN and to r$^2$SCAN-L
(PC\opt) and as well or better than OFR2 for solid test cases.  It
improves significantly on OFR2 for molecular test sets but still is
not as good for those as r$^2$SCAN-L(PCopt).  We leave until later
a more systematic approach to constructing smoothed functionals along
with a measure to assess the result.

Secondly, we report a study of timing characteristics, using a single
compute node.  We show that deorbitalization can be an attractive
option for meta-GGA calculations, reproducing similar if not better
results than original meta-GGA for solids and for equation of state
calculations, being up to twice as fast in total compute time as
gKS calculations.  At the same time, we find a number of
situations in which the two approaches are about equal in performance
and slow compared to GGAs, indicating need for future work.

In the remaining presentation, Section~\ref{sec:theory} details the
deorbitalization approach and the various KED models utilized. 
Sec.~\ref{sec:methods} describes the methods for performing structural and timing
benchmarks. Sec.~\ref{sec:results} reports the structural
bench-marking and detailed timing analysis for solids, including preliminary
\textit{ab initio} molecular dynamics (AIMD) outcomes. Sec.~\ref{sec:conclusion} presents
our conclusions and prospects for future work.

\section{Background \label{sec:theory}}
   \subsection{MetaGGA Structure \label{subsec:mgga}}

A conventional meta-GGA XC functional (ignoring spin-decomposition) is
defined by
\beq
E\xc[n] := \int d\bfr\ e\xc[n(\bfr),\nabla n(\bfr),\tau\s(\bfr)]
\label{eq:mggaconvdefn}
\eeq
where both the density $n$ and noninteracting kinetic energy density $\tau\s$ 
are expressed in terms of orbitals $\phi_i$ and occupation numbers $f_i$,
with
\beq
    n(\bfr)  :=  \sum_i f_i |\phi_i(\bfr)|^2 \; 
\label{eq:ndefn}
\eeq
and $\tau\s$ given by Eq.~(\ref{eq:tausdefn}).
To cope with the explicit orbital dependence 
in the XC energy (and therefore, implicit
density dependence), customarily the ground state energy is
found via the gKS procedure.  In it, the XC potential
is calculated as the 
functional derivative of the energy with respect to
the individual orbitals.   The result is the gKS equation
in which the kinetic energy operator becomes 
\beq
    -\frac{1}{2} \nabla^2 + 
            \nabla \cdot \left( \frac{\partial e\xc}{\partial \tau\s}\right)  \nabla \; ,
    \label{eq:gKSoperator}
\eeq
thus including a dependence upon the XC energy beyond that of 
the multiplicative potential of the pure KS formulation. 
(In practical implementation, that term can be rendered as an
orbital-dependent potential, hence still requiring 
a generalization of the KS procedure).

In SCAN and descendants such as r$^2$SCAN, 
$\tau\s$ enters the XC
functional via the iso-orbital indicator $\alpha$,
\beq
\alpha(n,\nabla n, \tau\s) := \frac{\tau\s-\tau\svW}{\tau\TF}  \;.
\label{eq:alphaKS}
\eeq
Here $\tau\svW$ is the von Weizs\"acker KED, namely that of a
system with a single occupied orbital and density $n$,
\beq
\tau\svW = \frac{1}{8} \frac{\left| \nabla n \right|^2 }{n}
\label{eq:tauwdefn}
\eeq
and $\tau\TF$ is the KED of the homogeneous electron gas of density $n$
\beq
    \tau\TF = \frac{3}{10} (3\pi^2)^{2/3} n^{5/3} \;.
\eeq 

Akin with other refinements of SCAN,  r$^2$SCAN uses
a regularized version of $\alpha$,
\beq
\bar{\alpha} := \frac{\tau\s-\tau\svW}{\tau\TF + 0.001\tau\svW }.
\label{eq:baralpha}
\eeq
This regularization is primarily to handle undesirable behavior for
exponentially small densities.  Nevertheless $\alpha$ serves as a
convenient point of reference from which to define orbital-free KE
density models.

The central roles of $\alpha$ and $\bar\alpha$ make it helpful 
to summarize, for context, the key structural elements of
r$^2$SCAN exchange.  Following Eq. (\ref{eq:mggaconvdefn}) with $e\xc
= n(\br)(\epsilon\x + \epsilon\c)$, for exchange we have
\bea
\epsilon\x^{\mathrm{r}{^2}\mathrm{SCAN}}(p, \bar{\alpha}) &=& \epsilon\x\LDA [n] F\x\rtwoSCAN (p,\bar{\alpha})  \\
\epsilon\x\LDA [n] &=& - (3/4)(3/\pi)^{1/3}n^{1/3} \\
\!\!F\x\rtwoSCAN (p,\bar{\alpha})&=&\lbrace  h\x^1(p) + f\x(\bar{\alpha}) %
\lbrack  h\x^0 - h\x^1(p) \rbrack \rbrace g\x(p) \;. \label{eq:r2scanelts}
\eea
Here, the dimensionless reduced gradient (squared) is
\beq
p := s^2  \equiv \frac{\vert \nabla n\vert^2}{4 (3\pi^2)^{2/3}n^{8/3}} \;.
\label{eq:pdefn}
\eeq
The $h\x^1(p)$ is a GGA form that is a rather complicated
function of $p$. It enforces the gradient expansion of exchange through
second order for $\alpha \approx 1$.
Switching between GGA forms for $\alpha \approx 1$ and the iso-orbital
limit ($\alpha \approx 0$), and extrapolation to $\alpha \to \infty$ is accomplished with the 
function $f\x(\bar{\alpha})$. It has two exponential regions ($\bar{\alpha} < 0$
and $\bar{\alpha} > 2.5$) joined by a seventh-order polynomial in $\bar\alpha$.   $g\x(p)$ is a damping function that goes
to unity in the limit $p \rightarrow 0$ and goes to zero as $p$ grows arbitrarily large.  Detailed expressions are in the Supplemental Material for
Ref. \citenum{FurnessEtAl2020}.

An important bound, given the centrality of $\alpha$, is 
\beq
    \tau\s \ge \tau\svW \; \Rightarrow \; \alpha \geq 0  \; .
\label{eq:alphapos}
    \eeq
This may be recognized as the non-negativity of the 
contribution to the Kohn-Sham kinetic energy derived from
Pauli exclusion. Note that the
r$^2$SCAN switching function $f_x(\bar{\alpha})$ just described is set up to enforce
this requirement even if numerical precision errors otherwise would
violate it.

The indicator $\alpha$ is
related trivially to the Pauli kinetic energy density which is the
subject of much of orbital-free DFT (OF-DFT)\cite{OFKEreview},
\beq
\tau_\theta := \tau\s - \tau\svW
\label{eq:tauthetadefn}
\eeq
or
\beq
  \alpha = \tau_\theta / \tau\TF \equiv F_\theta \; .
\label{eq:tautheta-Ftheta}
 \eeq
Here $F_\theta$ is the Pauli enhancement factor of OF-DFT.

To deorbitalize a conventional meta-GGA according to M-RT, the exact KS
KED is replaced by an approximate orbital-free (OF) semi-local
density functional
\beq
\tau\OF = \tau(n,\nabla n, \nabla^2 n) \;.
\label{eq:taumodel}
\eeq
The deorbitalized meta-GGA then is
\begin{multline}
E\xc[n] = \\ 
 \int \!e\xc\! \left\{ n(\bfr),\nabla n(\bfr),\tau\OF \!\left[ n(\bfr),
  \nabla n(\bfr), \nabla^2n(\bfr) \right] \right\} d\bfr \;.
\label{eq:deorbitalization}
\end{multline}
Minimization of the total energy with respect to $n$ then yields
a pure KS equation with a local XC potential of the form
%
\begin{multline}
v\xc = 
 \frac{\partial e\xc}{\partial n} 
   + \frac{\partial e\xc}{\partial \tau\OF} \frac{\partial \tau\OF}{\partial n}  - 
    \nabla \cdot\left( \frac{\partial e\xc}{\nabla n} + 
                      \frac{\partial e\xc}{\partial \tau\OF} \frac{\partial \tau\OF}{\partial \nabla n}\right)  \\ + 
    \nabla^2\left(\frac{\partial e\xc}{\partial \tau\OF} \frac{\partial \tau\OF}{\partial \nabla^2 n}\right) \;,
    \label{eq:vxc}
\end{multline}
In it, the first and third terms treat the explicit dependence of 
$e\xc$ on $n$ and $\nabla n$ respectively, while the others treat 
implicit dependence through $\tau\OF$.
A closely related quantity is the OF kinetic energy potential 
\beq
  v\s :=
       \left(\frac{\partial \tau\OF}{\partial n}\right) 
       - \nabla \cdot \left(\frac{\partial \tau\OF}{\partial\nabla n}\right)
       + \nabla^2 \left(\frac{\partial \tau\OF}{\partial\nabla^2 n}\right)  \;.
 \label{eq:vs}
\eeq
In the OF-DFT literature this is more commonly expressed as taking the functional derivative of a model Pauli KED, $\tau_\theta$. That yields the Pauli potential via 
\beq
v_\theta = v\s - v\svW,
\label{eq:vtheta}
\eeq
with $v\svW$ the KE potential for $\tau\svW$.

Comparison of Eqs.\ (\ref{eq:gKSoperator}) and\ (\ref{eq:vxc})
illuminates the distinct, key challenges for ordinary versus
deorbitalized meta-GGA XC numerical implementations.  A conventional,
orbital-dependent meta-GGA introduces a complicated kinetic energy
operator in addition to the ordinary local Kohn-Sham potential.
A deorbitalized counterpart produces a purely local
potential whose construction involves up to fourth spatial derivatives
of the density, with concomitant possible numerical difficulties.

\subsection{Deorbitalization models}

An orbital-free KED functional of the general form
Eq. (\ref{eq:taumodel}) obviously can be decomposed as in
Eq. (\ref{eq:tauthetadefn}). The result may be reduced by scaling
constraints to
\beq
\tau\OF(n,\nabla n, \nabla^2n) = F\SOF_\theta(p,q) \tau\TF(n) + \tau\svW(n,p).
\label{eq:taudeorbscaled}
\eeq
Here $F_\theta$ is as in Eq. (\ref{eq:tautheta-Ftheta}) and 
\beq
q  := \frac{\nabla^2 n }{4(3\pi^2)^{2/3} n^{5/3} }
\label{eq:qdefn}
\eeq
is the dimensionless reduced density Laplacian, partner to the 
reduced density gradient $p$ defined in Eq.~(\ref{eq:pdefn}).
Other important constraints include $\alpha$ non-negativity 
[Eq. (\ref{eq:alphapos})] and the gradient expansion of $F_\theta$ for slowly-varying 
density \cite{Kirzhnits57,Hodges}.

KED functionals that have proven particularly useful for
deorbitalization share several characteristics.  They are proper
meta-GGAs, in that they depend upon $n$, $p$, and $q$.  The design of
most has followed, to some degree at least, the strategy employed in
the PC KED functional~\cite{PerdewConstantin2007}, that is to build a
model for the KED in the slowly-varying density limit.  Then, because
models based on the formally correct gradient expansion in that limit
vary much more slowly with $p$ than the iso-orbital limiting case (the
von-Weizs\"acker functional, $\tau\svW$), eventually such models fail
the non-negativity constraint \cite{LevyOu-Yang1988}, $\tau\s \ge
\tau\svW$.  That failure is taken to indicate the onset of an
iso-orbital or nearly iso-orbital spatial region, and the slowly
varying KED is replaced by $\tau\svW$ by means of a suitable
switching function.

For PC and PC\opt~(the reparametrized PC used by M-RT
\cite{SCANL1,SCANL2}), the KED for the slowly-varying limit is a
modified fourth-order gradient expansion for which the Pauli 
enhancement function is 
\beq
F\SV\PC = \frac{1 +\Delta F_\theta^{(2)}+\Delta F_\theta^{(4)}}{\sqrt{1 + [\Delta F_\theta^{(4)}/(1 + 5p/3)]^2}}  \; .
\label{FthetaMGE4}
\eeq
The superscript ``SV'' denotes ``slowly varying''.  
The ingredient quantities are the second and fourth-order gradient
expansion corrections
\bea
\Delta F_\theta^{(2)}&:=& -\tfrac{40}{27} p + \tfrac{20}{9} q \\
\label{eq:gea2}
\Delta F_\theta^{(4)}&:=&\tfrac{8}{81} q^2 - \tfrac{1}{9} p q + \tfrac{8}{243} p^2  
\label{eq:gea4}
\eea
The second-order gradient correction $\Delta F_\theta^{(2)}$ consists of
a $5p/27$ contribution from the kinetic energy gradient expansion 
minus a $5p/3$ factor from the von Weizs\"acker KED
$(\tau\svW = (5p/3)\tau\TF)$.  That yields  an
overall negative slope with respect to $p$.  
Together with the possibility of negative $q$, this 
leads to eventual incipient violation of KED non-negativity which
forces the switch from
slowly-varying to the iso-orbital model form.

For PC, the interpolation function between the slowly-varying form 
of the Pauli enhancement factor and the
von Weizs\"acker lower bound $F_\theta\vW =0$ takes the form
\beq
\Theta\PC(x) = \begin{cases}
0,                                                               & x \leq 0 \\
f\PC(x/x_0)                                                    & 0 < x < x_0\\
1,                                                               & x \geq x_0 
\end{cases}
\label{eq:PCswitch}
\eeq 
where
\beq
f\PC(t) = \left[ \frac{1 + e^{1/(1-t)}}{e^{1/t} + e^{1/(1-t)}} \right]^b, \;\;   0 <  t < 1  \; . 
\label{eq:fpcdefn}
\eeq

Putting things together, the final form is
\beq
\alpha\PC = F_\theta^{\sss PC} = F\SV\PC \, \Theta\PC \left(F\SV\PC\right) \;,
\label{eq:PCform}
\eeq
with the reminder that $\alpha$ and the Pauli enhancement
factor $F_\theta$ are identical [Eq.~(\ref{eq:tautheta-Ftheta})]. 
The original PC parameter values are  $x_0 = 0.5389$, $b = 3$. 
The PC\opt~values are $x_0=1.784720$, $b=0.258304$ \cite{SCANL1},
determined by fitting to the KED $\alpha$ values of small-$Z$ atoms.  
That reparametrization helps to produce faithful deorbitalization
of some meta-GGAs but comes at the cost of a somewhat inaccurate form
for the slowly-varying limit ($p,q \to 0$).
 

Some drawbacks of the PC model were uncovered in Ref.~\onlinecite{CSK16},
which provided a modified form to fix them.  The switching
model $\Theta\PC$ was found to cause unphysical features in the KED in
covalent bonds, particularly for systems treated with pseudopotentials.
Concurrently, the PC limit for $q\to\infty$ and $p$ finite causes an
unphysical treatment of the exponentially decaying density in the
asymptotic region of a molecule, characterized by $q\to\infty$ and
$p/q \to 1$.  The first problem can be resolved by use of a switching
factor that obeys $\tau > \tau_{GE}$ and is as smooth as feasible.
The second problem can be resolved by recognizing that the
second-order gradient expansion for the KED has
the scaling behavior of the exact KED for an exponentially decaying
density and also is valid for small $p$ and $q$.  Thus it can suffice as a
bare-bones model for the slowly-varying limit.

Those choices, with further refinement~\cite{CR}, lead to the CR
model, namely
\beq
   \alpha\CR = 1 + \Delta F_\theta^{(2)} 
                    \Theta\CR\left(\Delta F_\theta^{(2)}\right)
\label{eq:alphacsk}
\eeq
with 
\beq
\label{eq:theta}
\Theta\CR(z) = \left[
        1 - \exp\left(-\frac{1}{|z|}^a
           \right)\left(1 - H(z)\right)\right ]^{1/a}  \; \; .
\eeq
$H(z)$ is the Heaviside unit step function.   
The exponent $a = 4.0$ produces reasonably close estimates
for the total KE of atoms\cite{CR}, while $a = 2.0$ produces somewhat smaller
Pauli potentials.

The RPP 
deorbitalizer
\cite{OFR2} is a PC variant designed specifically for
deorbitalizing r$^2$SCAN whilst retaining constraint
satisfaction.  As such it starts with the same form as Eq. (\ref{eq:PCform}), 
\beq \alpha\RPP = F\SV\RPP \Theta\RPP(F\SV\RPP)
\label{eq:RPPform}
\eeq
where $F\SV\RPP$ is, as before, a meta-GGA 
suitable for the slowly varying limit, to wit
%
%
%
\beq
F\SV\RPP =  1 + \Delta F_\theta^{(2)} + \Delta F_\theta'^{(4)} +
                \Delta F^{asy}_\theta.
\label{eq:Fsvrpp}
\eeq
In this expression, the fourth-order term $\Delta F_\theta'^{(4)}$
has the form of Eq.~(\ref{eq:gea4}) but with dramatically altered
coefficients: $b_{qq} = 1.801019$, $b_{pq} = −1.850497$, and $b_{pp} =
0.974002$.  Those correct a corresponding error in the r$^2$SCAN
exchange functional to restore gradient expansion compliance to fourth order
in the slowly-varying limit.  The last term of Eq.~(\ref{eq:Fsvrpp})
is higher than fourth-order and defines the asymptotic large $p,q$
behavior. It is given by
\beq
\begin{split}
\Delta F^{asy}_\theta & = c_3 p^2 \left(e^{-\vert c_3\vert p}-1\right) 
                       + \left(\Delta F_\theta'^{(4)} - c_3 p^2\right) \\
            & \times \left\{ \exp \left[ -\left(\frac{p}{c_1}\right)^2 - \left(\frac{q}{c_2}\right)^2\right] -1 \right\} \;.
\end{split}
\label{eq:deltaFtheta4}
\eeq
This imposes the
second-order gradient expansion as the limit for $p,q \to \infty$,
$p/q$ finite, as in CR. Optimal coefficient values were determined
against appropriate norms \cite{OFR2} to be
\bea c_1 &=& 0.202352\\
c_2 &=& 0.185020\\
c_3 &=& 1.53804  \;.
\label{eq:cvalues}
\eea

The switching functional $\Theta\RPP$ involves the same piecewise 
logic as in PC, Eq.~(\ref{eq:PCswitch}), but with the nonanalytic switching 
function $f\PC$, Eq.~(\ref{eq:fpcdefn}) replaced by the polynomial
\beq
f\RPP(t) = 20t^3 -45t^4 + 36t^5 -10t^6 \; .
\label{eq:RPPswitch}
\eeq
The switching constant $x_0 = 0.819411$. (There is no $b$ constant to set.)

In our context, a rather obvious step is to seek a smoothed RPP (SRPP)
functional by replacing the RPP switching functional with the CR form, 
\beq
\alpha\SRPP = 1 + (F\SV\RPP-1)\Theta\CR(F\SV\RPP-1)  \; .
\label{eq:SRPP}
\eeq
A somewhat smoother functional, denoted SRPP2, arises from employing
the exponent $a=2$ in $\Theta\CR$.

\begin{figure}[H]
  \includegraphics[angle=0,width=0.95\linewidth]{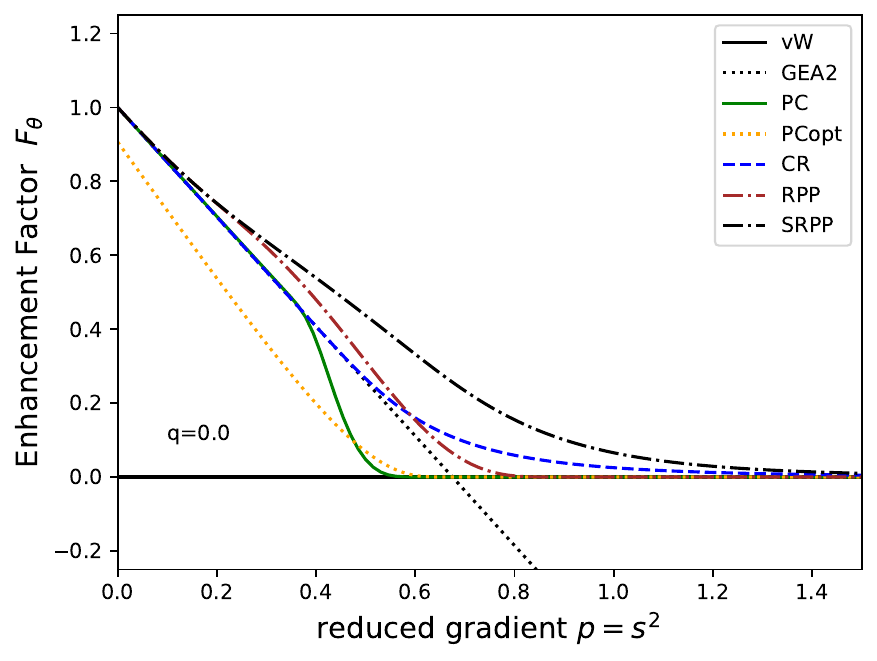}
  \centering
  \caption{Pauli enhancement factor $F_{\theta} (\equiv \alpha)$
for various KED functionals as a function of 
reduced density gradient $p$ with reduced Laplacian $q=0$. 
Models used are: GEA ($=1+\Delta F_\theta^{(2)}$), 
PC [Eqs.~(\ref{FthetaMGE4}-\ref{eq:PCform})], PC\opt (same form, altered
coefficients), 
CR [Eqs.~(\ref{eq:alphacsk}-\ref{eq:theta})], 
RPP [Eqs.~(\ref{eq:RPPform}-\ref{eq:RPPswitch})] and SRPP [Eq.~(\ref{eq:SRPP})].
$\alpha^{vW} = 0$ by construction. 
}
  \label{fig:enhance}
\end{figure}

Fig.~\ref{fig:enhance} shows how these design choices affect the
behavior of the Pauli enhancement factors, \textit{i.e.}, the $\alpha$
to be used in deorbitalization.  That figure presents $F_\theta$
versus $p$ for the specific value $q=0$.  The von-Weizs\"acker lower
bound value of $F_\theta\vW=0$ appears as a black solid line.  GEA2
denotes the gradient expansion model $1 + \Delta F_\theta^{(2)}$ that
characterizes the slowly varying limit, $p,q \to 0$.  The
slowly-varying limit for the PC meta-GGA starts at the homogeneous
electron gas value of 1 at $p=0$ and at first follows along the GEA2
trajectory.  But GEA2 transgresses the von-Weizs\"acker bound before
$p=0.8$.  (Recall the negative slope in the $p$-coefficient of the
gradient expansion discussed earlier.)  The switching function
$\Theta\PC$ avoids that crossing by inducing the sudden dip in PC off
the GEA2 onto the subsequent zero value, thus complying with the
non-negativity constraint for $F_\theta$.
That dip is a fairly clear candidate to suspect as the cause of
fluctuations in the PC Pauli potential and in the potential of a XC
functional deorbitalized with PC.  The dip leads to rapid variations
in the PC KED in regions of near iso-orbital density, such as covalent
bonds,~\cite{CSK16} which would be exacerbated in calculating the
potential.  This supposition is shown to be borne out in the following
section.  In contrast, the CR model (blue-dashed) and SRPP (black
dot-dashed) make the transition (from GEA2 to von Weizs\"acker bound)
in a much smoother fashion, while RPP is an intermediate case.
Finally $F_\theta$ for PC\opt~ fails to satisfy the
homogeneous electron gas limit ($F\to 1, p,q \to 0$) but does have a
reasonably smooth transition.

\subsection{Kinetic and exchange potentials \label{subsec:XfactorXpot}}

\begin{figure}[H]
  \includegraphics[angle=0,width=0.95\linewidth]{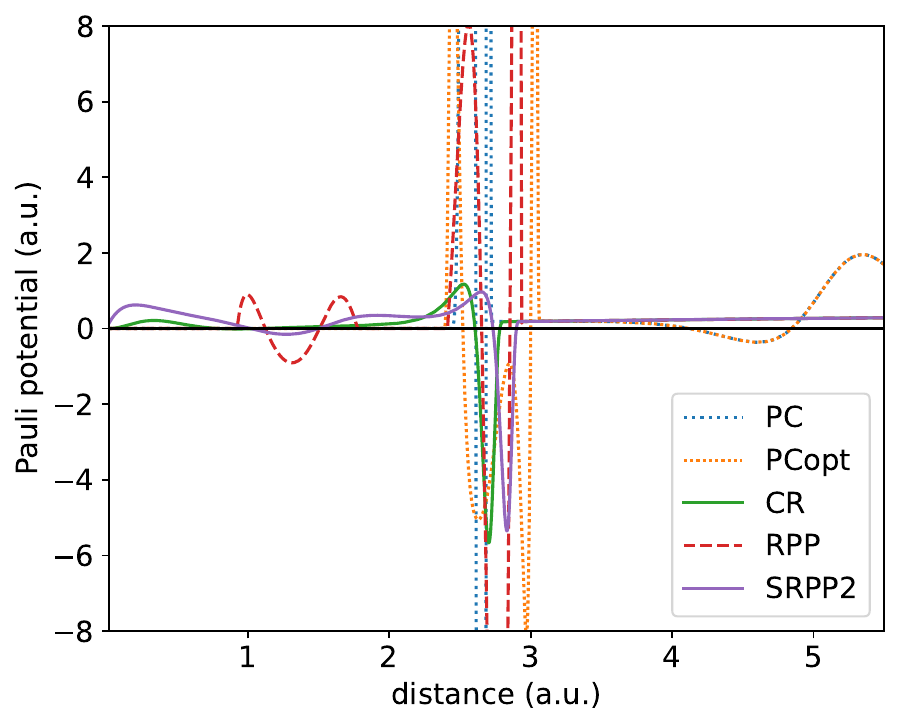}
  \centering
  \caption{Pauli kinetic potential $v_\theta = v_{\tau} - v\svW$ for various 
Laplacian-level KED functionals for the hydrogen atom, evaluated at the 
exact atomic density. The exact value for the H atom, $v_\theta=0$, is at the solid
black line.}
  \label{fig:pauli}
\end{figure}

The smoothing problems that arise in the deorbitalizers are
well-illustrated by a plot of the kinetic potential $v_\theta = \delta
T_\theta/\delta n$, Eq. (\ref{eq:vtheta}), for various relevant
Laplacian-dependent KED functionals for the Hydrogen atom, evaluated with the exact
H atom density.  See Fig.~\ref{fig:pauli}.
Most of the approximate KED functionals deliver potentials that are
reasonably close to the exact value (zero) except in the vicinity of
2.5 to 3.0 a.u. There violent fluctuations in the potential are
evident. Those can be traced to a spurious diagnosis of density
behavior. At high densities, the KED functionals diagnose hydrogen to
be an iso-orbital system, which is correct.  But at low density they
treat it like a slowly varying electron gas.  
The fluctuations occur for the values of $r$ where the transition between 
the two behaviors occurs.
The problem is worst for
PC which suffers oscillations larger than 100 hartrees. The problematic
behavior is
somewhat less severe for PC\opt~and RPP with oscillations in the tens
of hartrees.  Only the CR model~\cite{CR} and its extension to the RPP
form, SRPP2, show more reasonable variations, on the order of a few
hartrees.  In fact, the size of those variations is correlated closely
(with the exception of PC\opt) with the abruptness of the function
used to describe the transition (from iso-orbital to slowly varying
electron gas) in the enhancement factor, as seen in
Fig.~\ref{fig:enhance}.  As a caution, we note that the plot shows
potentials evaluated with the exact ground-state density for
hydrogen. The unreasonable oscillatory behavior almost certainly would
be worse for self-consistent densities from the associated potentials.

\begin{figure*}[ht]
   \vfill
   \begin{subfigure}[b]{0.49\textwidth}
       \includegraphics[angle=0,clip=true,scale=0.27]{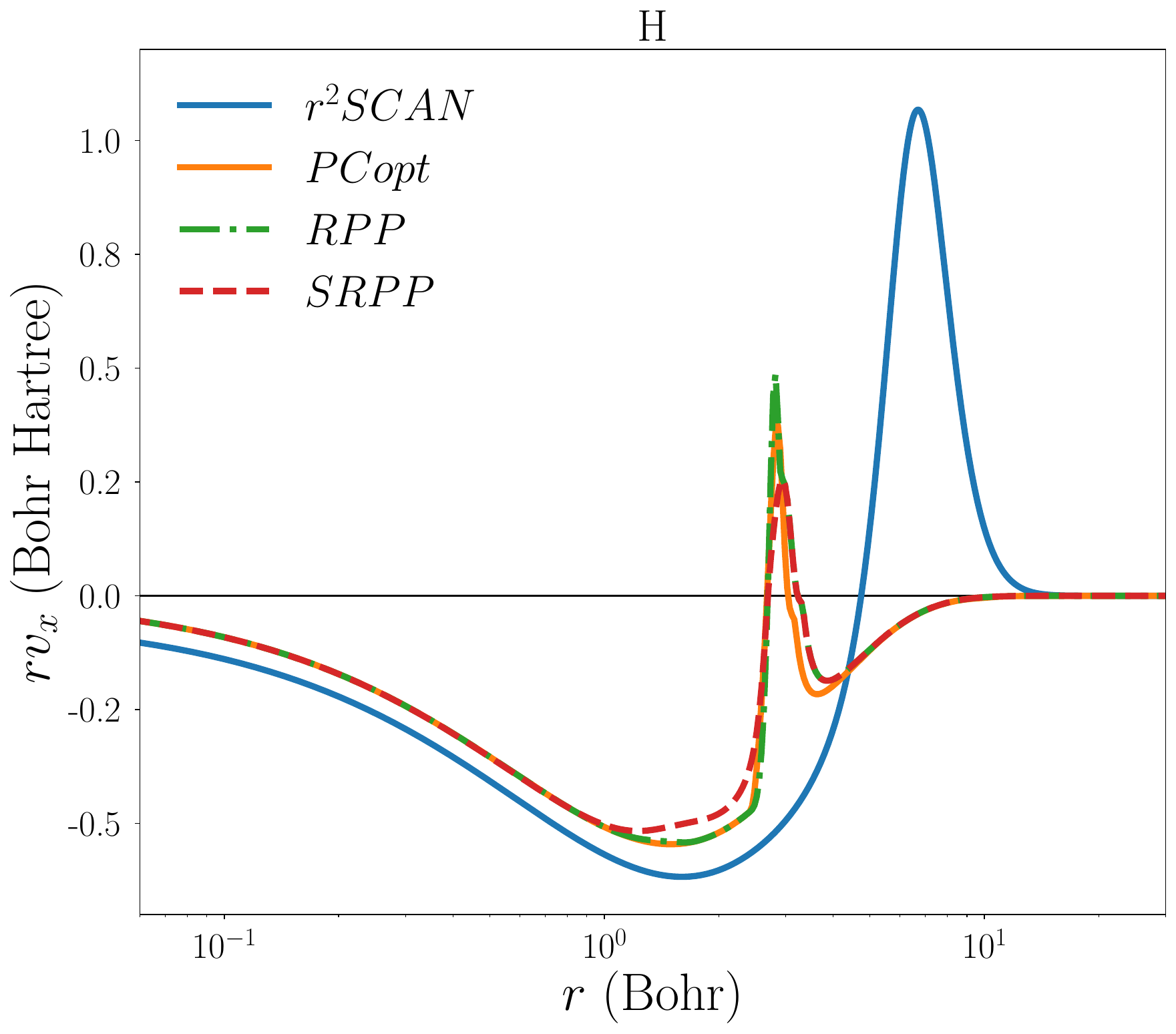}
       \centering
   \end{subfigure}
   \hfill
   \begin{subfigure}[b]{0.49\textwidth}
       \includegraphics[angle=0,clip=true,scale=0.27]{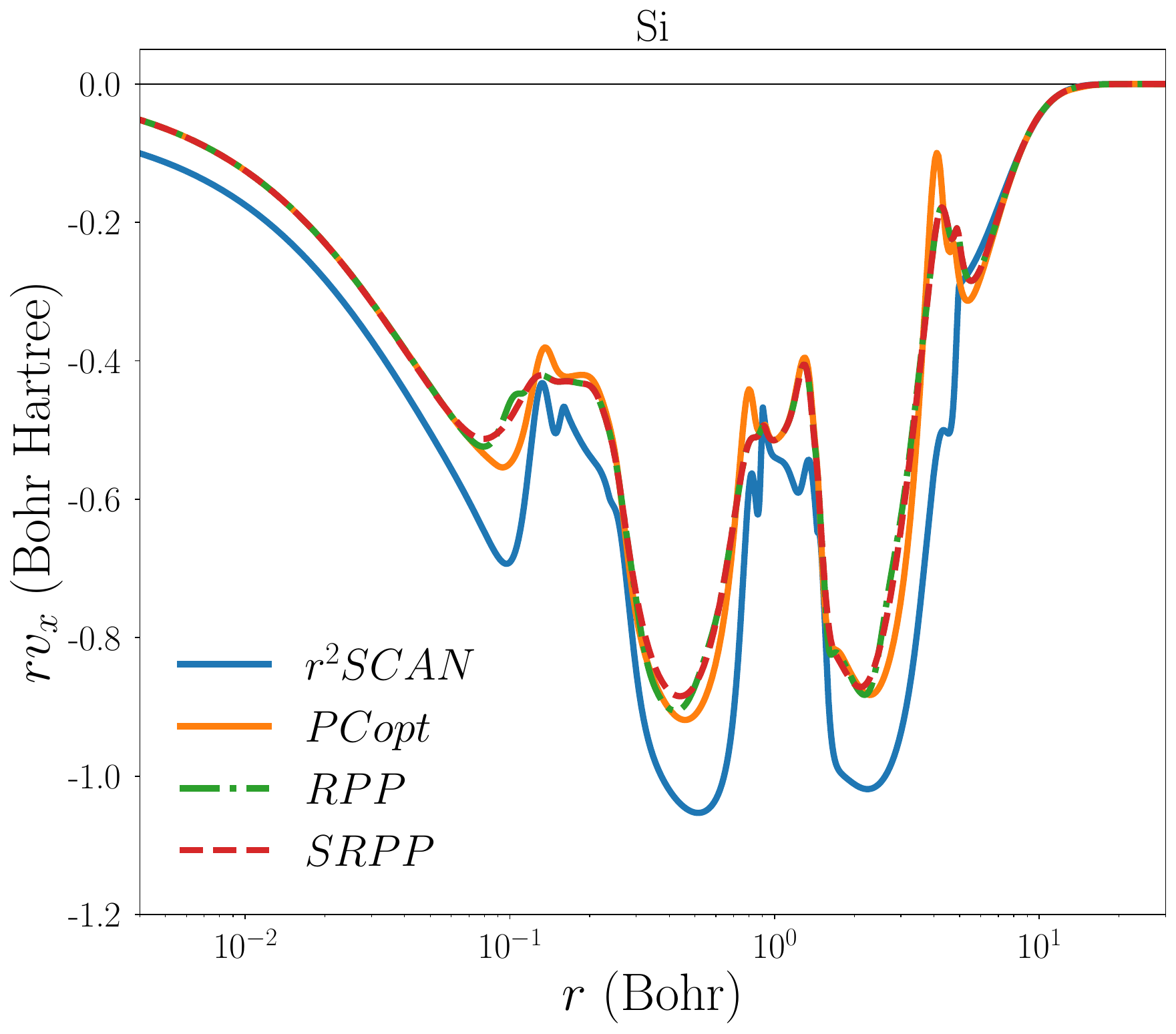}
       \centering
   \end{subfigure}
   
\caption{Local part of the r$^2$SCAN exchange potential for the H and Si atoms
 (scaled by radius $r$)
 compared to the exchange potentials of several deorbitalized versions of r$^2$SCAN for the same atoms.}
\label{fig:xc_pot1}
\end{figure*}

\begin{figure*}[ht]
\includegraphics[angle=0,clip=true,scale=0.27]{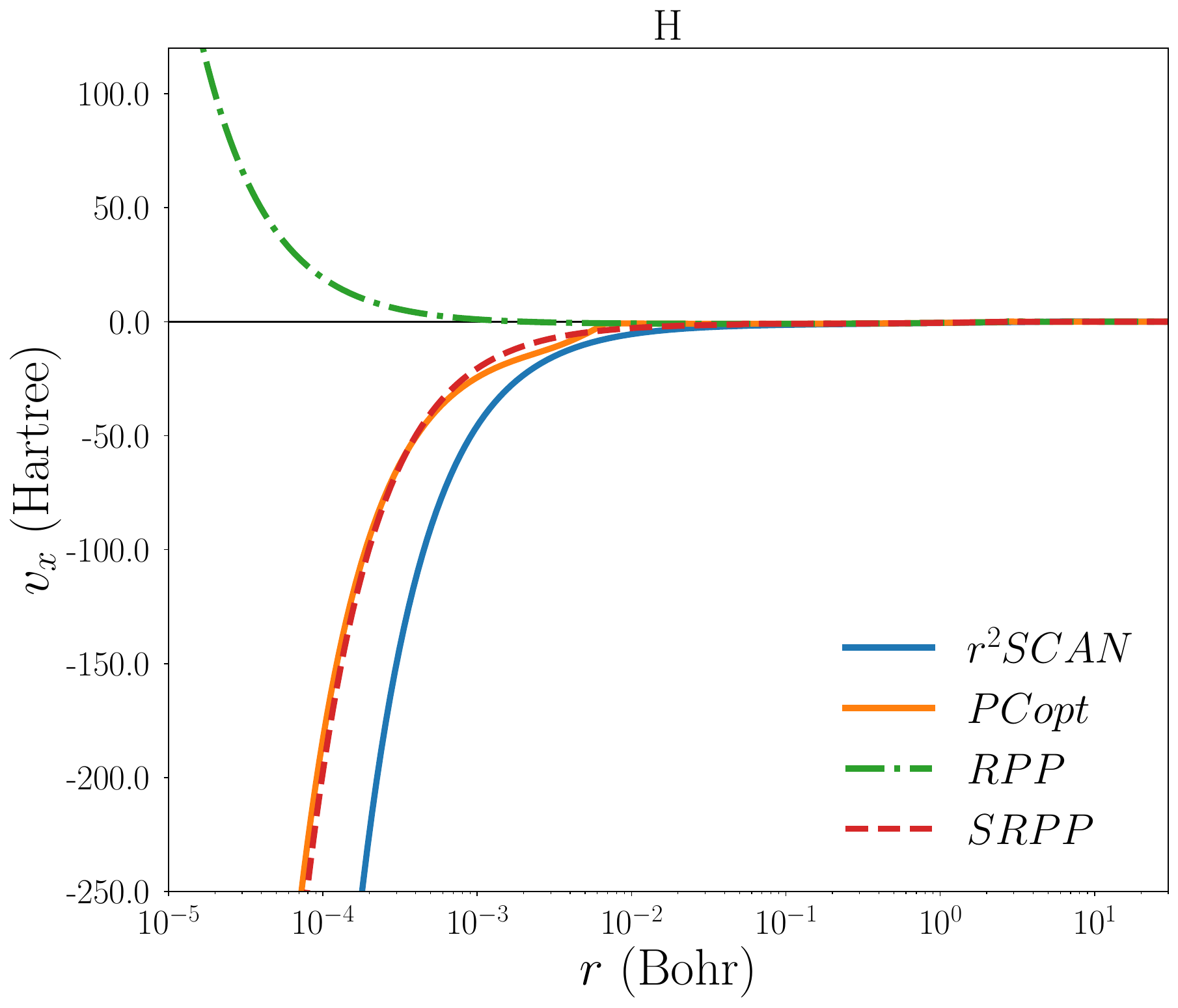}
\centering
\caption{
The local part of the r$^2$SCAN X potential for atomic H compared to deorbitalized
X potentials, plotted over a much larger range in length and energy.}
\label{fig:xc_pot2}
\end{figure*}

Figures \ref{fig:xc_pot1} and \ref{fig:xc_pot2} show plots of the
local part of the exchange potential (Eq.~\ref{eq:vxc}) for the parent
meta-GGA functional (r$^2$SCAN) and the full X potential for three
deorbitalized versions of it (PC\opt, RPP, and SRPP). The systems are
the H atom (at two different length scales) and the Si atom.  (Aside:
r$^2$SCAN deorbitalized with the RPP form has the formal functional
name OFR2. 
As a generalizable name for any deorbitalized meta-GGA that is reasonably
  compatible with previous usage, e.g. M-RT, from here on we have adopted  
the portmanteau of meta-GGA,  ``-L" (for Laplacian dependence), and deorbitalizer name in parentheses.  
Our new deorbitalizations of r$^2$SCAN thus are r$^2$SCAN-L(SRPP) and (SRPP2).
For simplicity in making comparisons between  
deorbitalization options, where context allows, frequently we will refer to
the deorbitalizer name,  not the full functional name.)
Note that the X potential in Fig.~\ref{fig:xc_pot1} is multiplied by
the radial coordinate to bring out features near the valence edge
where most variation in potentials occurs. For r$^2$SCAN, the
contribution from the functional derivative with respect to the
kinetic energy density is omitted, as it is incorporated in the
physics of the gKS scheme via Eq.~(\ref{eq:gKSoperator}).  These
potentials are computed using a modified version of the Python code
\textit{densities}~\cite{densities} 
which employs 
fixed analytic spherical densities obtained from self-consistent Hartree-Fock calculations.~\cite{HFdensitiesA} 
They were evaluated on a double-exponential grid with 1000 radial points, 
with substantially more points for r$^2$2SCAN.

First, we note a comparison with a prior calculation of the 
r$^2$SCAN-L(RPP)
(OFR2) X potential as a function of radial distance for the Hydrogen atom,
Fig.~7 of Ref.~\citenum{OFR2}.  Even allowing for how the
information is displayed, [$rV\x(r)$ on the $y$ axis here versus
$V\x(r)$ in the other paper and $r$ on a log scale here versus uniform
in the other)], it is clear that the two noise profiles are profoundly
different. Ours shows considerably less noise, assuming any sharp
feature in the potential outside the cusp region to be noise
induced. As shown in the Supplemental Information \cite{SuppInfo}, we have analyzed
the dependence of the numerically calculated X potential upon grid
density. We find oscillations of magnitude similar to those shown 
in Ref.~\citenum{OFR2} when our grid has 300 points.  Those vanish
with finer grids and converge to the value shown here for a grid size
greater than 700 points. We also verify that the two potentials,
when plotted in the same fashion, are essentially identical up
to noise.  We do not know if the prior work included a noise
sensitivity analysis, but doing so seems to be an important consideration
when working with the density Laplacian. 

Though an unequivocal comparison of the gKS local part and ordinary KS
potentials is not possible, it is clear that deorbitalization of
r$^2$SCAN introduces unphysical oscillations.  In particular, we see
large noise in the region $r \approx 3$~au for the H atom, where the
Pauli potential, Fig.~\ref{fig:pauli}, is seen to fail.  
However, the
SRPP deorbitalization of r$^2$SCAN [r$^2$SCAN-L(SRPP)] results in 
modestly less noise in
the potential compared to the PC\opt~or RPP deorbitalizations.  A
similar trend occurs for Si, with SRPP slightly smoother than RPP and
both smoother than PC\opt.  Interestingly, the RPP and SRPP produce
potentials noticeably smoother overall than the local part of the
parent meta-GGA, despite use of the density Laplacian.  
(To be sure that this indeed is true, we needed to increase the
resolution for the r$^2$SCAN results to 10,000 points for H and
40,000 for Si.)

The noticeable bump in the r$^2$SCAN  H atom potential at large $r$
[Fig.~\ref{fig:xc_pot1}(a)] may be a remnant of a
major issue in the original SCAN.  In a region of exponentially
decaying density, $v_x$ for SCAN goes to a large positive value rather
than zero \cite{BartokYates19,FurnessEtAl2020}. That unphysical
behavior forces the use of extremely fine grids and leads to
difficulties in obtaining pseudopotentials.\cite{PAWS22} (Note that
the exact KS exchange potential for the H atom must cancel the
self-Hartree potential, hence $rv_x$ must go asymptotically to $-1$,
which no potential considered here does properly.)

The short-range behavior for any atom is exemplified by the H atom
case.  All potentials for it fail by diverging at extremely small
distances, rather than trending to a small negative
value~\cite{LB94}. Overall $v_x$ from PC\opt~and SRPP are rather close.
Unlike $v_x$ from RPP, they are roughly similar quantitatively
to the parent $v_x$. 

   \subsection{Noise problem quantification}

The unphysically noisy features in the Pauli kinetic potentials shown in
Fig.~\ref{fig:pauli}  can be diagnosed readily as arising primarily from the density Laplacian term, i.e., the last term of Eq.~(\ref{eq:vtheta}). Clearly
it is closely related to the last term in Eq.~(\ref{eq:vxc}) for the
exchange-correlation potential. This
makes sense since both have the Laplacian of a functional of the
Laplacian of the density.  Thus fourth spatial derivatives of the density arise
in both as well as 
third derivatives of $e\xc$ in the latter case. Those high-order
spatial derivatives intrinsically 
accentuate oscillations in a function.  Reduction of the magnitude
of this term should lead to smoother
potentials.  The reduction in $v_\theta$ oscillations from PC to CR
shown in Fig.~\ref{fig:pauli} is commensurate with that
supposition.  Such reduction should yield more efficient numerical performance. 

For a quantitative assessment of the extent to which that supposition
is correct, we borrow a measure from electrostatics.  Observe that 
satisfaction of Laplace's equation
\beq \nabla^2 \psi = 0
\eeq
by some function $\psi$ within a specified spatial region  is the condition for
minimizing the action defined by
\beq
I = \frac{1}{2} \int |\nabla \psi|^2 dV 
\eeq
with respect to variations in $\psi$, with the value
of $\psi$ on the boundary of the solution region fixed.

The impact of the Laplacian term in the kinetic potential thus can be measured 
by calculating the action for an appropriate $\psi$.  
Here we choose
\beq
 I^{deorb} = \frac{1}{2} \int \left | \nabla 
     \left ( \frac{\partial \tau}{\partial \nabla^2 n} \right )
     \right|^2 dV  \;.
    \label{eq:ideorb}
    \eeq
This technique has been applied previously to minimize
the Laplacian term of the overall exchange-correlation potential -- 
see Ref.~\onlinecite{CWW12} for details.
Our choice to concentrate on the kinetic potential reflects the fact 
that the Laplacian term of the X
potential actually generates two plausibly problematic contributions.
One, $\nabla^2 (\partial e\xc/\partial \tau)$, arises from the
functional dependence of $e\xc$ on $\tau$.  The other, $\nabla^2
(\partial \tau/\partial \nabla^2 n)$, arises from the dependence of
the deorbitalization of $\tau$ on $\nabla^2 n$.  The effects of the
form of $e\xc(\tau)$ upon performance characteristics of both
conventional and deorbitalized meta-GGAs has been the subject of
recent attention.~\cite{BartokYates19,FurnessEtAl2020, OFR2} However,
comparison of Fig.~\ref{fig:pauli} to Fig.~\ref{fig:xc_pot1} shows
that the location of major error in the hydrogen X potential (the bump
in deorbitalized potentials around 2 to 3~au) coincides with that
of dramatic noise in the Pauli potential.  This implicates, strongly, 
the second term as a culprit in generating X potential noise, hence
that term is a focus of concern.
A second benefit of this choice is to focus on an aspect of
the potential that is independent of the
choice of XC functional being deorbitalized.

Minimization of oscillations in that term on a useful test
density or densities would have the potential of minimizing its impact on
numerical performance in general. Perhaps that should be a
design goal for a deorbitalized functional. We will explore that
idea in a follow-up to the present study.  For now, to get an
efficient, useful estimate of the impact of deorbitalizer-generated
noise, we simply evaluate Eq.~(\ref{eq:ideorb}) for the H atom as
calculated from a given KED functional.  The crucial point is that the
exact Pauli potential, Eq.~(\ref{eq:vtheta}), for this case is zero.

The consequences of that evaluation are shown, along with the kinetic
energy obtained by integrating $\tau\OF$, in
Table~\ref{Table:Ideorb} with the deorbitalizers PC, PC\opt, CR, RPP, SRPP,
and SRPP2. There is a close correlation between the smoothness of 
transition between slowly-varying and iso-orbital limits exhibited by
the enhancement factors seen in Fig.~\ref{fig:enhance} and the
quantified measure $I^{deorb}$.  The switching behavior of PC clearly is
the most abrupt. Thus it has an enormous $I^{deorb}$ value. 
PC\opt~ and RPP do much better, with a reduction in $I^{deorb}$ by
a factor of five and an order of magnitude respectively.  
Clearly, the functionals with the CR switch (CR, SRPP, SRPP2) are the 
best.  That pattern is manifest to a considerable extent in the
corresponding kinetic potentials (Fig.~\ref{fig:pauli}) 
and to a
lesser but still significant extent for X potentials, Fig.~\ref{fig:xc_pot1}.

The next issue, obviously, is to address to what extent (if any) this
smoothing endeavor and its quantitative assessment has consequences for
calculations on real systems.  

\begin{table*}[ht]
\centering
\caption{Kinetic energy and integrated noise measure $I^{deorb}$
evaluated on the Hydrogen atom for several deorbitalized models of the 
KE density.}
\renewcommand{\tabcolsep}{6pt}
\begin{tabular}{@{}lccccccc@{}}
\toprule
\hline
Model & Exact & PC & PCopt & RPP & CR & SRPP & SRPP2 \\
\midrule
KE (ha) & 1/2 & 0.507 & 0.506 & 0.504 & 0.514 & 0.527 & 0.550 \\
I (au) & 0    & 220   & 44.0 & 19.0 & 1.705 & 1.755 & 1.555 \\
\hline
\bottomrule
\end{tabular}
\label{Table:Ideorb}
\end{table*}
       
\section{Computational methods \label{sec:methods}}

For molecular system tests, calculations were done with the
NWChem 7.0.2 code \cite{NW} using the def2-TZVPP basis set and
\texttt{xfine} grid settings. Test comparisons are to heats of
formation computed according to the now-standard procedure of Curtiss
\textit{et al.}~\cite{G31,G32} for the 223 molecules of the G3X/99
test, to  bond lengths for the T96-R test set~\cite{DF1,DF2}, and
to harmonic vibrational frequencies using the T82-F test
set~\cite{DF1,DF2}.

Solid structural properties were calculated with the VASP-5.4.4
code~\cite{VASP}, following the same methodology and protocol as in
Ref.~\citenum{srTMpartI}.  

In particular, we used projector augmented wave (PAW) methodology with
PBE-based datasets, consistent with previous M-RT studies
\cite{SCANL2}.  (Since PAWs for meta-GGAs are unavailable in VASP, use
of PBE PAW datasets for meta-GGAS in VASP calculations is accepted
practice.)  
For hexagonal close-packed structures the ideal $c/a$
ratio was used.  For cohesive energies, isolated atom energies were
calculated using a $14 \times 15 \times 16 \AA$ unit cell and
$\Gamma$-point sampling.  Static-crystal lattice constants and
cohesive energies are compared to published results for 55 solids
\cite{55s} and, correspondingly, for bulk moduli of 44 solids
\cite{44s}. In addition, band gaps of 21 insulators and semiconductors
\cite{21s} were computed.  Equilibrium lattice constants $a_{0}$ and
bulk moduli $B_{0}$ at $T=0^{\circ}K$ were determined by calculating
the total energy per unit cell at twelve points in the volume range
$V_{0} \pm 10 \% $, where $V_{0}$ is the calculated equilibrium unit
cell volume, followed by fitting to the stabilized jellium equation of
state (SJEOS)~\cite{SJEOS}.  All the error values are from comparison
with tabulated experimental values.

For timing calculations, the same methodology was used except that for
the 55 solids \cite{55s} test set, the calculations were redone at the
experimental lattice constants.  In that way only single-point energy
evaluations were needed for timings.  Calculations used a single node
on the University of Florida HiperGator system (Gen.\ 3) with an AMD EPYC 75F3
32-core processor with 4GB of memory per core. They were executed in
parallel across all 32 cores.

\section{Results \label{sec:results}} 
\subsection{Atoms}

An issue raised by moving from the RPP to SRPP is norm compliance.
The free coefficients $c_1$, $c_2$, and $c_3$, are determined in the
RPP by fitting to 1) the XC energy of noble gas atoms Ne, Ar, Kr, Xe;
2) ``cluster" energies of jellium clusters; and 3) surface energies of
jellium clusters.
In developing the SRPP with the goal of smoothing the potential relative
to RPP, we have not modified any 
aspect of the RPP except for the 
transition between the slowly varying and iso-orbital limits. 
The $c$ coefficients and other parameters in the slowly-varying form of RPP are left 
unchanged.
Thus, it would be unsurprising if the norms fit by them 
were to be less well met in SRPP than in RPP.

We can test the effect that replacing the piecewise polynomial
transition function [Eqs.~(\ref{eq:PCswitch}) and
  (\ref{eq:RPPswitch})] in RPP with the smoother SRPP function
[Eq.~(\ref{eq:theta})] has on satisfaction of the ``appropriate norm"
of atomic exchange-correlation energies.
We calculated self-consistent RPP and SRPP exchange-correlation
energies for atoms up through $Z=18$ using NWChem, with the same basis sets as for molecules. 
As benchmark energies, we take highly accurate correlation
energies from Ref.~\citenum{Davidson93} and 
exact exchange calculation values using
the optimized effective potential code OPMKS~\cite{ED99}.
 
Overall, we find a mean error of about 2.1\% for the RPP and
2.8\% for SRPP, both systematically negative.  This error decreases
gradually in magnitude with system size.  That decrease is expected since the
XC energy of large-$Z$ atoms is increasingly well-characterized by the
slowly-varying electron gas as $Z \to \infty$~\cite{BCGP16}, and each
functional matches that limit as a norm.  The discrepancy between RPP
and SRPP for atoms between $Z=10$ and $18$ is about 0.5\%, decreasing rapidly
with $Z$.  This rapid decrease is consistent with the
decrease in importance of the cusp region and the exponential 
decay of the density far from the nucleus, the two regions in which 
the transition from gradient expansion to von Weizs\"acker behavior is most 
important. See the Supplemental Information for more details.

We lack the capacity to test compliance with the jellium norms, but
expect similar results.

\subsection{Structural properties\label{subsec:structure}}

Evidently the comparison of primary interest is for r$^2$SCAN versus
its deorbitalized versions, r$^2$SCAN-L(PC\opt) , with the
PC\opt~deorbitalizer
\cite{r2SCANL}, as well as with r$^2$SCAN-L(RPP) \cite{OFR2} and
with r$^2$SCAN-L(SRPP) and (SRPP2). An underlying issue, discussed in detail above,
is that PC\opt~breaks some constraints that the other three
preserve.  That distinction motivates further comparison, namely
between PC\opt~and the three more constraint-compliant deorbitalizers.

Tables \ref{Table:mol_struct} and \ref{Table:sol_struct} compile mean
errors (ME) and mean absolute deviations (MADs) for r$^2$SCAN, and
deorbitalized r$^2$SCAN, for the four deorbitalizers, along with
results from PBE \cite{PBE} as a baseline. In addition to those
standard measures of functional performance, we include the spread of
errors as defined by the maximum (most positive) error minus the
minimum (most negative.) This last measure is motivated by the wide
variations in timing performance that we discuss below.  That
variability raises the question whether there might be similar
variability in predictive performance.

\begin{table*}
\caption{Comparison of molecular structural property results for 
  the r$^2$SCAN XC functional and various deorbitalized-(r$^2$SCAN-L)
  variants derived from PC\opt, RPP, SRPP, and SRPP2 KED functionals.
PBE results included for context.  Heat of formation mean errors (ME), mean absolute deviation (MADs), and spread 
in kcal/mol, bond length errors in \AA, and frequency errors in cm$^{-1}$.}

\renewcommand{\tabcolsep}{12pt}
\begin{tabular}{@{}llccccccc@{}}
\toprule
     &      &\multirow{2}{*}{PBE}  &\multirow{2}{*}{r$^2$SCAN}   &\multicolumn{4}{c}{r$^2$SCAN-L} \\
     &      & & & \textit{PC\opt}~& \textit{RPP} & \textit{SRPP} & \textit{SRPP2}   \\
\midrule
\hline
\multirow{3}{*}{Heats of Formation} 
       & ME  & -20.878 & -3.145 & 1.845  & 8.796  & -5.918  & -9.470    \\
       & MAD & 21.385 & 4.488  & 5.300  & 13.109 & 7.804  & 10.405 \\
       & Spread & 88.091 & 29.818 & 42.938 & 72.117 & 48.329 & 46.036\\
       &        &        &        &        &        &        &        \\
\multirow{3}{*}{Bonds}
       & ME & 0.018 & 0.005 & 0.008  & 0.014  & 0.011  & 0.013   \\
       & MAD & 0.018 & 0.010 &0.011  & 0.014  & 0.012   &0.014  \\
       & Spread & 0.168 & 0.183 & 0.197 & 0.182 & 0.069 & 0.180 \\
       &        &        &        &        &        &        &        \\
\multirow{3}{*}{Frequencies} 
       & ME  & -33.781 & 11.336  & -7.248 & -26.743  & -22.875 & -22.945\\
       & MAD & 43.613 & 30.899  & 25.709 & 36.711   & 36.134  & 36.800 \\
       & Spread & 261.16 & 212.42 & 201.73 & 266.59 & 276.09 & 268.84 \\
\hline 
 \bottomrule
\end{tabular}
\label{Table:mol_struct}
\end{table*}

\begin{table*}[ht]
\centering
\caption{As in Table~\ref{Table:mol_struct} for solid-state structural properties.  Mean errors
  (MEs), mean absolute deviations (MADs) and spreads for equilibrium
  lattice constants in \AA, cohesive energies in eV/atom, bulk
  moduli in GPa, and band gaps in eV.}
\renewcommand{\tabcolsep}{12pt}
\begin{tabular}{@{}llcccccc@{}}
\toprule
     &      &\multirow{2}{*}{PBE}&\multirow{2}{*}{r$^2$SCAN}&\multicolumn{4}{c}{r$^2$SCAN-L} \\
     &      & & & $PC_{opt}$ & \textit{RPP} & \textit{SRPP} & \textit{SRPP2}   \\
\midrule
\hline
\multirow{2}{*}{Lattice constants} & ME  & 0.046 & 0.026  & 0.022  & 0.003  & -0.003 & -0.002 \\
                                   & MAD & 0.053 & 0.037  & 0.038  & 0.029  &  0.028 &  0.028 \\
                                   & Spread & 0.222 & 0.311 &0.212 & 0.154  & 0.143  & 0.142  \\
                                   &     & &        &        &        &        &        \\
\multirow{2}{*}{Cohesive energies} & ME  & -0.070 & -0.134 & -0.327 & -0.017 &  0.051 &  0.064 \\
                                   & MAD & 0.252 & 0.238  & 0.349  & 0.217  &  0.224 &  0.227 \\
                                   & Spread & 2.096 & 2.343 & 1.841 & 1.726 & 1.773  & 1.801 \\
                                   &     & &        &        &        &        &        \\
\multirow{2}{*}{Bulk modulus}      & ME  & 9.704 & 1.367  & -4.249 & 1.084  &  1.928 &  2.228 \\
                                   & MAD & 11.022 & 5.963  & 10.115 & 8.542  &  7.866 &  7.851 \\
                                   & Spread & 64.09 & 63.67 & 96.90 & 81.64 &  76.25 & 76.25  \\
                                   &     & &       &        &        &        &        \\
\multirow{2}{*}{Band Gaps}         & ME  & -1.69 &-1.20  & -1.38  & -1.60  & -1.58  &  -1.57 \\
                                   & MAD & 1.69 &1.20   & 1.38   & 1.60   & 1.58   &   1.57 \\
                                   & Spread & 4.89 & 4.32 & 4.72  & 4.99   & 4.93   & 4.92 \\

\hline 
 \bottomrule
\end{tabular}
\label{Table:sol_struct}
\end{table*}

For molecular systems, the results presented in Table~\ref{Table:mol_struct} 
show that SRPP is a substantial improvement
(40\% reduction) over RPP for heat of formation MAD.  SRPP2 also
provides an improvement over RPP, albeit a significantly smaller one.  
On bond lengths and
vibrational frequencies, the three (RPP, SRPP, SRPP2) perform about the
same. None proves competitive with PC\opt.  
All three RPP-derived deorbitalizations show a disappointing 
degradation of performance, especially for heats of formation and frequencies, 
compared to PC\opt~deorbitalization, and none can be considered
a completely faithful deorbitalization of r$^2$SCAN.
Also note the near 
equality of the magnitude of MEs and MADs for SRPP2 and to a lesser
extent, SRPP.  This, and
the fact that the MEs for both are negative, indicate that they
tend to underbind consistently, leading to larger MADs.

For solids, Table~\ref{Table:sol_struct} shows quite different
outcomes.  SRPP, SRPP2, and RPP all have distinctly better MADs than
PC\opt~for lattice constants, cohesive energies, and bulk
moduli. SRPP and SRPP2 are a tiny bit better than RPP on lattice constant MAD,
slightly worse (3\%) on cohesive energy, and about 9\% better on bulk
modulus.  The bulk modulus spread for SRPP and SRPP2 is better than
for RPP as well and substantially better than that from PCopt~.

Overall, SRPP and SRPP2 preserve the formal properties of RPP and
either improve on its performance or maintain it.  RPP is already
known to be better on metallic solids than PC\opt, (and better in 
some measures than r$^2$SCAN) but worse on
molecules \cite{OFR2}. That behavior is confirmed by our results
(again, see Tables \ref{Table:mol_struct} and \ref{Table:sol_struct}).
Among the three, SRPP and SRPP2 are slightly better than RPP for solids,
while SRPP is the best of the three over all systems though not competitive with PC\opt~for
molecules.  These outcomes reinforce the finding by Kaplan and Perdew
\cite{OFR2} that re-introduction of compliance with constraints broken
by PC\opt~actually can reduce the breadth of applicability of the
deorbitalization.

The error spread reduction provided by the RPP-derived models was
unexpected. Those spreads are quite a bit better than the original
r$^2$SCAN for both cohesive energies and lattice constants and not
much worse for bulk moduli.  In that regard,  PC\opt~shows some degradation.
Similar results are found if we measure standard deviations.  That is
to say, deorbitalization can, in some cases, lead to a significant
reduction in outliers compared to the gKS calculations with the parent
meta-GGA XC functional, an unexpected bonus.  The inverse correlation
of spread to our measure of potential smoothness is suggestive. If
some of that spread is due to numerical instability rather than
functional accuracy, these results would be a trade-off between
removing numerical problems with gKS and introducing problems with the
Laplacian.  The relative absence of outliers leads us to speculate
that the noise reduction in smoothed deorbitalized potentials may
reduce the risk that geometry optimization procedures will discover
spurious local minima.

\subsection{Timing results \label{subsec:times}}

Table~\ref{Table:moltime} presents timing statistics for the 223
molecule G3X/99 test set.  Table~\ref{Table:ae6time} displays the
timing data for the six-molecule subset, denoted AE6, of the G3 test
set. Each table shows the total time for the run of a test set, and
the average time taken per system.  In addition, each table shows the
average number of SCF cycles needed to converge to self-consistency,
(defined as the total number of SCF cycles divided by the number of
systems) and the average time per SCF cycle, taken as the average time
divided by the average number of cycles.  In
Table~\ref{Table:moltime}, spreads in these last two measures are
shown also.  As before, spreads are defined as the difference between
the maximum and minimum values of a quantity observed across a data
set.

Table~\ref{Table:moltime} shows that both r$^2$SCAN-L(SRPP) and (SRPP2) are
substantially faster than r$^2$SCAN-L(RPP) for the molecules. Similar to the
PC\opt~ deorbitalization, neither of them delivers a speed advantage
over the parent, orbital-dependent functional.  The effects of
test-set sampling are shown in Table~\ref{Table:ae6time}.  The AE6
timings give a much more favorable comparison of SRPP2 over RPP but no
meaningful gain for SRPP2 versus either PC\opt~ or the parent
r$^2$SCAN functional.

Methodological effects are shown in Table~\ref{Table:G3_vasp}.  It
gives the timings for the G3 test with the VASP 5.4.4 calculations
done as isolated systems in a large orthorhombic box. Distinct from
the solid calculations, the default cutoff energy was set to $600$
\textit{eV} and $\Gamma$ point sampling was used.  Unlike the G3
calculations in NWChem, the VASP tests exhibit a large degradation in
performance by both RPP and SRPP deorbitalizations with respect to the
parent functional. Interestingly, the
time per SCF cycle for the SRPP deorbitalizer is 1.29 s/cycle in VASP,
while in NWChem it is 1.28 s/cycle. However, for the parent functional
r$^2$SCAN, VASP takes 2.87 s/cycle compared to 1.34 s/cycle in
NWChem. Unfortunately, the number of SCF cycles is much higher for the
deorbitalized forms, roughly a factor of 2.5 for SRPP and a factor of
3 for RPP. In NWChem the average number of SCF cycles is roughly the
same for all three.  The speed advantage per cycle of the
deorbitalized forms is lost thereby in the quasi-molecular case.  This
brings to light the drastic effect that basis set methodology (plane-wave
PAW versus gaussian) can have on the performance of the deorbitalized
forms.

\begin{table*}[ht]
\centering
\caption{Timings and number of SCF cycles required for computing the 223 molecules of the G3X/99 test set. Time measurements are expressed in seconds ($s$).}
\renewcommand{\tabcolsep}{6pt}
\begin{tabular}{@{}lcccccccc@{}}
\toprule
\hline
\multirow{2}{*}{223 molecules} &\multirow{2}{*}{r$^2$SCAN}&\multicolumn{4}{c}{r$^2$SCAN-L}& \multirow{2}{*}{PBE} \\
           & & \textit{PC}\opt~ & \textit{RPP} & \textit{SRPP} & \textit{SRPP2}   \\
\midrule 
Total time                & 2432.00 & 2410.90 & 3036.90 & 2433.60  & 2373.70  & 1417.70 \\
Average time              & 10.91    & 10.81    & 13.62    & 10.91    & 10.64    & 6.36 \\
                          &          &          &          &          &          &  \\
Avg. number SCF cycles         & 7.79     & 8.07     & 8.42     & 8.19     & 8.28     & 8.45  \\
Spread                    &23 &24  & 24 &24  &24 & 22 \\
                          &          &          &          &          &          &  \\
Avg. time per SCF cycle & 1.34     & 1.28     & 1.56     & 1.28     & 1.23     & 0.72  \\
Spread                    &6.27 & 6.25 & 7.38 & 6.14 & 6.13 & 4.10 \\  
\hline
\bottomrule
\end{tabular}
\label{Table:moltime}
\end{table*}

\begin{table*}[ht]
\centering
\caption{As in Table~\ref{Table:moltime} for the AE6 test set.}
\renewcommand{\tabcolsep}{12pt}
\begin{tabular}{@{}lccccccc@{}}
\toprule
\hline
\multirow{2}{*}{AE6} &\multirow{2}{*}{r$^2$SCAN}&\multicolumn{4}{c}{r$^2$SCAN-L} \\
           & & \textit{PC}\opt~ & \textit{RPP} & \textit{SRPP} & \textit{SRPP2}   \\
\midrule 
Total time                & 25.40 & 25.40 & 35.00& 27.20  & 26.00 &  \\
Average time              & 4.23  & 4.23  & 5.83  & 4.53   & 4.33  &  \\
                          &       &       &       &        &       &  \\
Avg. number SCF cycles         & 6.50  & 6.33  & 8.33  & 7.17   & 7.17  &  \\
                          &       &       &       &        &       &  \\
Avg. time per SCF cycle & 0.66  & 0.67  & 0.76  & 0.65   & 0.62  &  \\

\hline
\bottomrule
\end{tabular}
\label{Table:ae6time}
\end{table*}

\begin{table*}[ht]
\centering
\caption{As in Table~\ref{Table:moltime} for the 223 molecules, but with the calculations done in VASP 5.4.4 for the functionals deorbitalized with RPP and SRPP as compared to the  parent 
functional.}
\renewcommand{\tabcolsep}{12pt}
\begin{tabular}{@{}lccccccc@{}}
\toprule
\multirow{2}{*}{223 molecules} &\multirow{2}{*}{r$^2$SCAN}&\multicolumn{2}{c}{r$^2$SCAN-L} \\
           && \textit{RPP} & \textit{SRPP}  \\
\midrule 
Total time               &11100.38 & 13573.66 & 13030.01 \\
Average time             &49.78 & 60.87    & 58.43    \\
                         & &          &          \\
Average SCF cycles        &17.22 & 51.90    & 43.14    \\
                          &&          &          \\
Average time per SCF cycle& 2.87& 1.10     & 1.29     \\
\hline
 \bottomrule
\end{tabular}
\label{Table:G3_vasp}
\end{table*}


\begin{table*}[ht]
\centering
\caption{Timings and number of SCF cycles required for computing the 55 solids. Times are expressed in seconds ($s$).}
\renewcommand{\tabcolsep}{10pt}
\begin{tabular}{@{}lcccccccc@{}}
\toprule
\hline
\multirow{2}{*}{55 solids} &\multirow{2}{*}{r$^2$SCAN}&\multicolumn{4}{c}{r$^2$SCAN-L} & \multirow{2}{*}{PBE}\\
           & & \textit{PC}\opt~ & \textit{RPP} & \textit{SRPP} & \textit{SRPP2}   \\
\midrule 
Total time                & 6052.81 & 7812.30 & 6024.63 & 4892.41 & 4553.67  & 2650.44  \\
Average time              & 110.05  & 142.04  & 109.54  & 88.95  & 82.79  & 48.19 \\
                          &          &          &          &          &          &  \\
Avg. number SCF cycles         & 15.64   & 60.31   & 40.35   & 30.49  & 31.42   & 14.20 \\
Spread                     & 32 & 341 & 242 & 176 & 138 & 19 \\
                          &          &          &          &          &          &  \\
Avg. time per SCF cycle & 6.93    & 2.56    & 2.91    & 2.97   & 2.89   &  3.26\\
Spread                     & 22.39 & 7.87 & 7.30 & 6.97 & 7.46& 7.77 \\
\hline
\bottomrule
\end{tabular}
\label{Table:soltime}
\end{table*}

For solids, Table~\ref{Table:soltime} presents a clearly different story. 
r$^2$SCAN deorbitalized with SRPP or SRPP2 outperforms the parent,
orbital-dependent functional on total time. SRPP2 in particular needs
only about 75\% of the time of r$^2$SCAN.  Notably, SRPP2 also
requires only about 72\% more time than PBE.  This overall gain
compared to the PC\opt~ deorbitalization comes from a drastic
reduction in the number of SCF cycles needed by SRPP and SRPP2, about
half of the number required by PC\opt.  Importantly, the times per SCF
cycle for SRPP and SRPP2 are essentially the same as for RPP, less
than half that for r$^2$SCAN, and actually faster than even for
PBE. The combined result is an overall speedup.

The required number of cycles follows, reasonably closely, the measure
of smoothness in the Laplacian contributions in the exchange
potential (Table~\ref{Table:Ideorb}), hence strongly implicates those 
terms in the slow SCF
convergence performance seen in some of the deorbitalizations.  We
surmise that the sensitivity of such terms to rather small changes in
the density makes achievement of a self-consistent density harder than
in gKS or GGA calculations.  This diagnosis is supported in the
extreme spread of numbers of cycles for deorbitalized functionals
compared to the parent functional as evaluated via gKS.  The spread in
observed cycle counts for r$^2$SCAN (gKS) is only slightly more than
twice the average number of cycles, while for the deorbitalized forms
(KS), the spreads range from 4.4 to 6.0, with SRPP2 the best
performer.

\begin{figure}[ht]
   \begin{subfigure}[b]{0.49\textwidth}
       \includegraphics[angle=0,clip=true,width=0.95\linewidth]{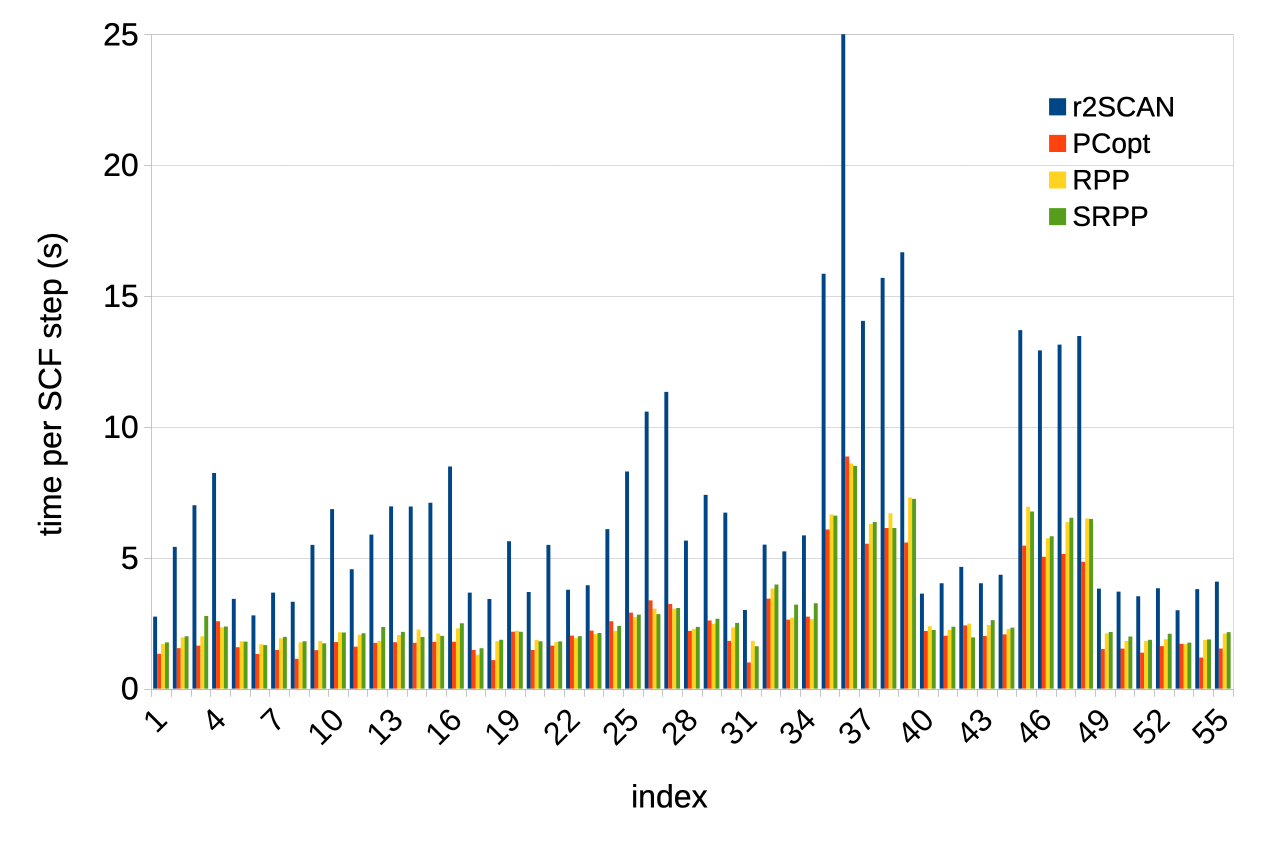}
       \centering
       \caption{Average time per SCF cycle in seconds.}
       \label{fig:timeperstep}
   \end{subfigure}
   
   \begin{subfigure}[b]{0.49\textwidth}
     \includegraphics[angle=0,clip=true,width=0.95\linewidth]{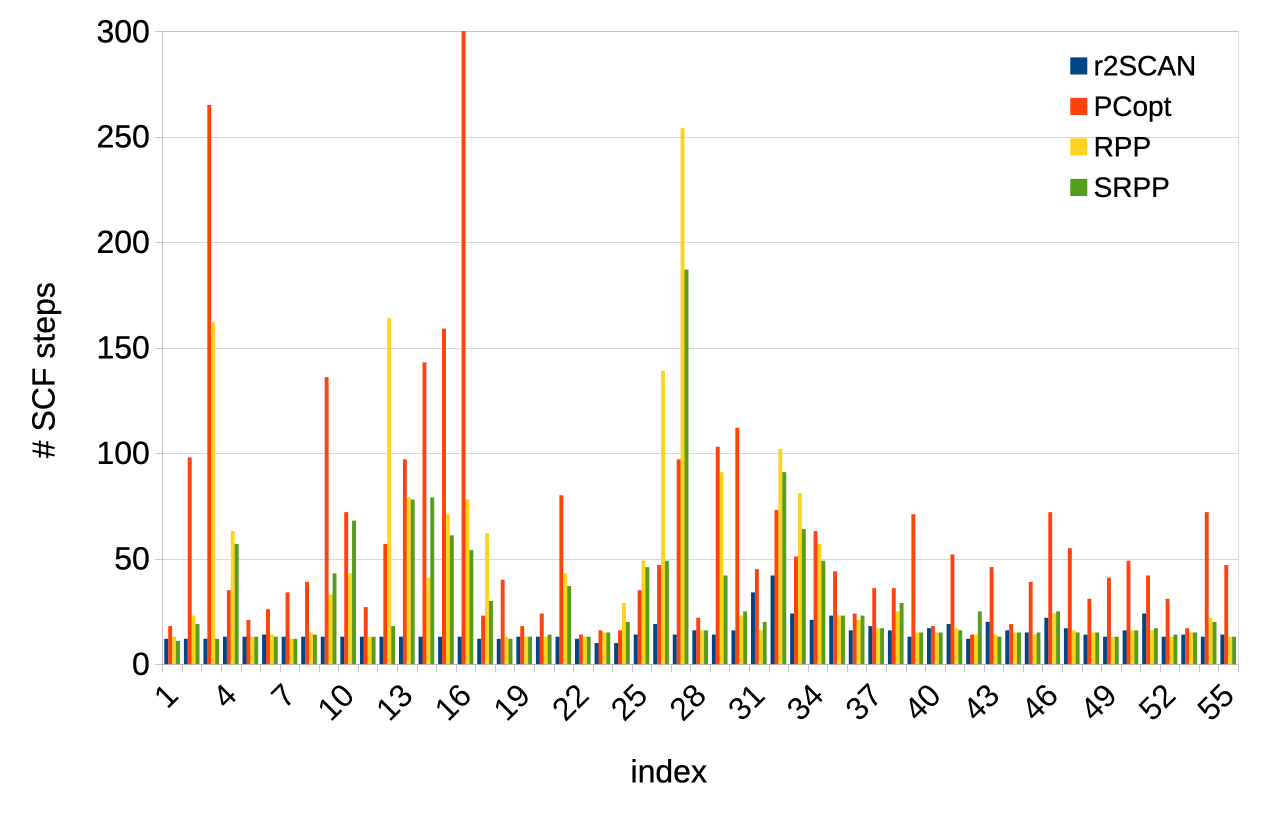}
     \centering
     \caption{Number of SCF cycles.}
     \label{fig:barchart}
   \end{subfigure}
\caption{Bar charts showing timing performance
for r$^2$SCAN, and r$^2$SCAN-L with  the PC\opt, RPP, and SRPP deorbitalizers for the entries in the 55-solid test set.  
Indexing of materials: 1-4 elemental semiconductors; 5-16: compound semiconductors; 
17-22: ionic compounds; 23-31: simple metals; 32-55 transition metals.
}
\end{figure}

More insight can be found by visualizing the timings for the
individual members of the solid test set. These are shown as average
time per SCF cycle in Fig.~\ref{fig:timeperstep} and number of SCF
cycles in Fig.~\ref{fig:barchart}.  Results for r$^2$SCAN itself as
well as versions deorbitalized with PC\opt, with RPP, and with SRPP are shown.
The distribution for SRPP2 is only modestly better than SRPP, 
so it was omitted for clarity.

\begin{figure}[ht]
   \begin{subfigure}[b]{0.49\textwidth}
       \includegraphics[angle=0,clip=true,scale=0.32]{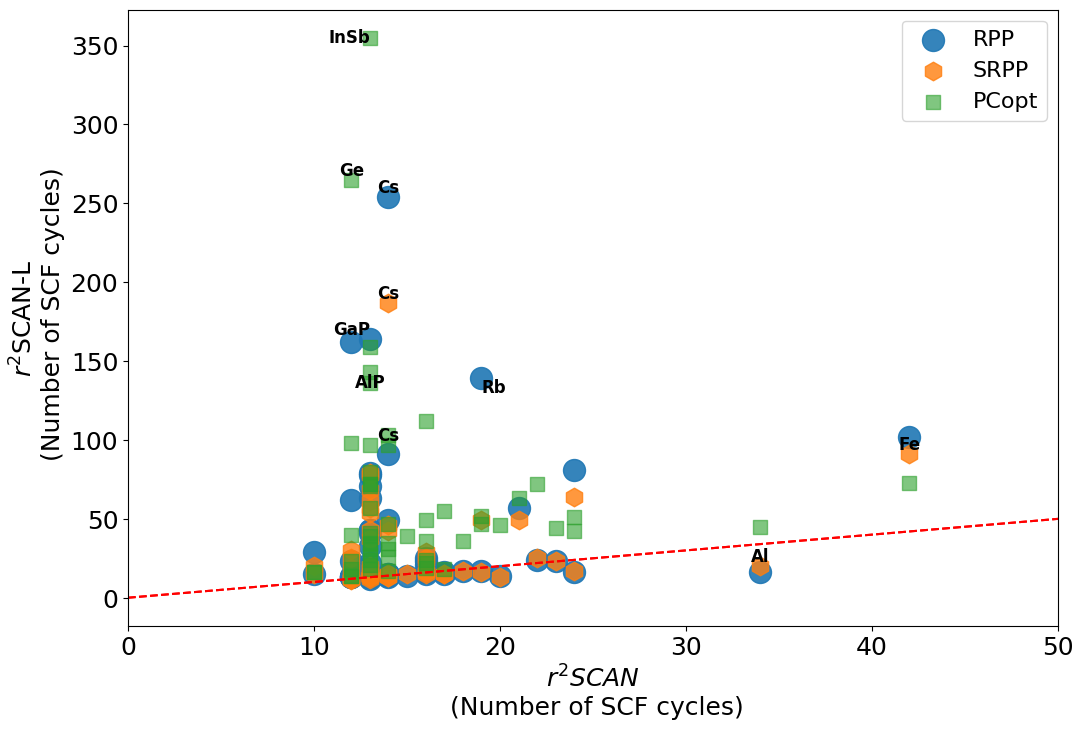}
       \centering
   \end{subfigure}
\caption{Scatter plot showing the number of SCF cycles for 55 solids
  calculated using the deorbitalized functionals relative to the
  number of cycles required with the r$^2$SCAN functional. The red
  dotted line is the locus of values for the
  deorbitalized functionals that are equal to the r$^2$SCAN
  values.}
\label{fig:scatter_scf_cycles}
\end{figure}

Fig.~\ref{fig:timeperstep} shows the clear SCF cycle-time advantage of
the deorbitalized functionals versus their orbital-dependent parent,
including the nearly 3 times faster performance of PC\opt~ originally
reported by M-RT.  The other deorbitalizers are slightly less swift
but still well above twice as fast as r$^2$SCAN per cycle.  Notably,
the issue is one of outliers. For about half the data set, the
r$^2$SCAN (gKS) time is under 5~s and not quite two times slower than
the deorbitalized forms (with KS potentials).  But there are multiple
bad actors in the orbital-dependent case. For them the slowdown is
much worse.  Deorbitalized functionals do poorly on a much smaller
data set of nine transition metals, all with nearly empty \textit{d}
shells, e.g. Y or Sc, or nearly empty spin subshells, like Fe.

Fig.~\ref{fig:barchart} shows the number of SCF cycles needed for
r$^2$SCAN and the three deorbitalized versions considered in the
preceding plot.  One sees clearly how the SCF process for original
r$^2$SCAN converges in consistently fewer cycles than for any of the
three deorbitalized variants. The RPP-type deorbitalizers perform
quite reasonably but with a number of outliers. In contrast PC\opt~ has
a rather larger number of outliers which cancel its edge in the
time-per-cycles metric. Note also the scatter plot of SCF cycles needed
versus the r$^2$SCAN requirement in Fig.~\ref{fig:scatter_scf_cycles}.
The smoothed r$^2$SCAN-L(SRPP) and (SRPP2) do provide consistent
improvement upon the performance of r$^2$SCAN-L(RPP), with a few exceptions among the
semiconductors and simple metals.

In sum, while there will be remaining issues in achieving
self-consistency with Laplacian-dependent functionals, it is clear
that such issues are ameliorated significantly by having a smoother
potential.

\subsection{Detailed analysis of computational parameter dependence}


Given the importance of materials calculations with plane-wave basis
codes such as VASP and the significant performance differences between
molecules and solids for such codes, investigation of computational
technique and parameter choice effects is imperative.  We pursue that
in two ways. In this subsection we examine the dependence of timing
results on various run parameter choices.  In the following subsection
we consider the equation of state (cold curve) of Al and a short 
AIMD calculation for it.

Table~\ref{Table:changes_all} provides timing information for
r$^2$SCAN versus r$^2$SCAN-L deorbitalized via RPP and SRPP for the 55
solid test set. It shows the effects of differing choices of the
energy cutoff, convergence parameter, and minimization algorithm.
Cases labeled ``0'' through ``3'' successively reduce the convergence
tolerances while keeping the SCF minimization procedure fixed as the
preconditioned conjugate gradient method. That is the recommended
option for meta-GGA exchange-correlation functionals.  Cases 4, 5, and
6 explore effects of the use of less-expensive diagonalization
algorithms.

\begin{table*}[ht]
\centering
\caption{Parameter and SCF algorithm dependence of timing for computing the set of 55 solids, for r$^2$SCAN itself, and  r$^2$SCAN-L deorbitalized with RPP or SRPP. $E_{\mathrm{cut}}$ is the plane-wave cut-off energy; $E_{\mathrm{diff}}$ is the energy tolerance used to
end the self-consistency cycle.  The algorithms to solve the KS or gKS equation,
using the keyword terminology for the ALGO input parameter in VASP, are:
(All) preconditioned conjugate
gradient, (Normal) blocked Davidson iteration, (Fast) hybrid of block-Davidson and 
RMM-DIIS (Residual Minimization Method with Direct Inversion in the Iterative Subspace), and (Very Fast) RMM-DIIS.
}
\renewcommand{\tabcolsep}{6pt}
\begin{tabular}{@{}lccccccccc@{}}
\toprule
         & Case 0 & Case 1& Case 2 & Case 3 & Case 4 & Case 5 & Case 6\\
\midrule
Algorithm       &All     &All     &	All    & All     &Normal    & Fast & Very Fast\\
$E_{\mathrm{diff}}$ (eV) &$1\times10^{-6}$ &$2.72\times10^{-5}$&$1\times10^{-6}$ & $2.72\times10^{-5}$ &$1\times10^{-6}$ & $1\times10^{-6}$ & $1\times10^{-4}$ \\
$E_{\mathrm{cut}}$ (eV) &800              &800                   & 600             & 600                    &800             &800               &500\\
\hline 
\textit{Average time}    & & & & & & \\
r$^2$SCAN             & 110.05  & 101.28  & 95.51   & 81.47   & 112.26  & 104.26 & 58.37 \\
RPP             & 109.54  & 63.78   & 55.23   & 42.02   & 103.97  & 63.11 & 22.66 \\
SRPP            & 88.95    & 59.11   & 55.56   & 45.56   & 111.93  & 79.25 & 23.62 \\

\textit{Average SCF cycles}       & & & & & &         \\
r$^2$SCAN            & 15.64   & 12.67   & 15.78   & 12.78   & 12.91    & 12.93 & 17.36 \\
RPP             & 40.35   & 20.93   & 20.11   & 13.75   & 29.96   & 27.33 & 18.51 \\
SRPP            & 30.49    & 17.55   & 17.49   & 13.69   & 29.09   & 30.80 & 18.75 \\

\textit{Average time/cycle}   & & & & & &         \\
r$^2$SCAN    & 6.93    & 7.74    & 5.96    & 6.16    & 8.59    & 7.91 & 3.34 \\
RPP             & 2.91    & 3.14    & 2.77    & 2.98    & 3.73    & 2.45 & 1.21 \\
SRPP            & 2.97     & 3.32    & 3.03    & 3.21    & 4.11    & 2.83 & 1.24 \\
\hline
 \bottomrule
\end{tabular}
\label{Table:changes_all}
\end{table*}

Note that for all cases there is a significant advantage for the
deorbitalized forms with respect to the parent functional for the time
taken by a single calculation of electronic orbitals.  The speed-up
stays a little over a factor of two, maybe better for the fastest
eigensolver.  

In contrast, the advantage in terms of the number of cycles needed to
reach self-consistency for the gKS procedure used with r$^2$SCAN over
the deorbitalized forms and their pure KS potential does vary quite a
bit depending on the tightness of convergence criteria. 
gKS loses its advantage relative to deorbitalized KS as convergence is
loosened and SRPP loses its advantage relative to RPP. 
Ultimately, for
Case 3, the sloppiest one for preconditioned conjugate 
gradients and Case 6, the sloppiest overall,
there is basically no difference in
performance among the three strategies in this regard.  
Therefore
the total time speed-up of deorbitalization is quite dramatic, almost
a factor of 2.
(It may seem peculiar that all functionals in Case 6
take substantially more cycles to achieve self-consistency than Case
3, with both larger plane-wave cutoff and stricter convergence
tolerance.  This is a result of the instability of the RMM-DIIS, which
makes sacrifices in stability to achieve faster overall times \cite{VASPworkshop}.

There is a plausible, simple explanation for such a
finding. Reducing the cut-off energy amounts to introducing a low-pass
filter that eliminates a good deal of noise in the potential.  Though
the r$^2$SCAN-L(RPP) and (SRPP) potentials have very different noise characteristics,
it is reasonable to suppose that those differences are substantially
removed by smoothing with a low energy cutoff.  Similarly, a less
demanding tolerance means greater acceptance of effects of noise in
the potential, reducing their consequences.  We hypothesize that the
smooth atomic-like basis used in molecular calculations is coarse
relative to individual plane waves, therefore causing a similar noise
reduction in the potential and yielding the ``egalitarian" timing
results seen for these calculations.

\subsection{Molecular dynamics of Aluminum \label{sec:MD}}

For context, in Table~\ref{Table:altime} we provide the timing and
SCF count for calculating the static lattice fcc Al equation of
state at zero temperature.  The data are averages over runs at
12 lattice parameters, 3.8484~\AA  $\le a_0 \le$ 4.1146~\AA. (At each lattice
constant, the calculation was started from the same density, not
from the equilibrium density of the preceding lattice constant.
This technical aspect is of importance later.)

Solid Al is a particularly advantageous case for
deorbitalization. Consistent with the  discussion above about the 55
solid test set (recall Table~\ref{Table:soltime}), all three
deorbitalizations (PC\opt, RPP, SRPP) are substantially faster in
total time than the parent functional.  The spreads are even better
for r$^2$SCAN-L(RPP) and (PC\opt), and as good as the parent for (SRPP).  Also as
expected, the RPP- and SRPP-deorbitalized functionals  
outperform PC\opt~substantially with regard to the required number of SCF cycles.

\begin{table*}[ht]
\centering
\caption{Timings (in seconds, $s$) and number of SCF cycles required for computing the zero-temperature fcc Al 
  static lattice equation of state for r$^2$SCAN and three deorbitalizations, with PBE values for context.}
\renewcommand{\tabcolsep}{6pt}
\begin{tabular}{@{}lcccccccc@{}}
\toprule
\hline
\multirow{2}{*}{fcc Al} &\multirow{2}{*}{r$^2$SCAN}&\multicolumn{3}{c}{r$^2$SCAN-L}& \multirow{2}{*}{PBE} \\
           & & \textit{PC}\opt~ & \textit{RPP} & \textit{SRPP}  \\
\midrule
Total time              & 990.89 & 611.87 & 366.93 & 415.89 & 296.04 \\
Average time            & 82.57  & 50.99  & 30.58  & 34.66  & 24.67  \\
                        &        &        &        &        &        \\
Avg. number SCF cycles  & 26.75  & 45.00  & 17.58  & 20.83  & 12.42  \\
spread                  & 23.00  & 51.00  & 9.00   & 19.00  & 4.00   \\
                        &        &        &        &        &        \\
Avg. time per SCF cycle & 3.12   & 1.16   & 1.75   & 1.68   & 2.00   \\
spread                  & 0.42   & 0.32   & 0.21   & 0.41   & 0.30   \\
\hline
\bottomrule
\end{tabular}
\label{Table:altime}
\end{table*}

Those rather encouraging timing and cycle count results would lead to
the expectation that the SRPP and RPP functionals would outperform the
parent in driving AIMD.  To investigate that,
we obtained timing statistics for a short AIMD simulation of
liquid-phase aluminum.

The simulation system consisted of 108 atoms, at bulk density
$\rho=2.34~\mathrm{g/cm}^{3}$ in a $12.7239 \times 12.7239 \times
12.7239$~\AA$^3$ cell. The $\Gamma$-point was used to sample the
Brillouin zone. In VASP 5.4.4, the system was treated with a
three-electron PAW effective potential ($3s^2$ $3p^1$ valence, 10 electrons in
the core; denoted PAW\_PBE Al\_GW). Non-spherical contributions within
the PAW spheres were included self-consistently.  Minimization used
the RMM-DISS algorithm with energy cutoff set to 500~eV, as in Case 6
in Table~\ref{Table:changes_all}.  The self-consistent energy
convergence tolerance $E_{\mathrm{diff}}$ was $1 \times 10^{-4}$~eV.
Approximate Fermi-type thermal smearing with a width of 0.0881555~eV
was applied, and the initial temperature set to
1023~K. 
The molecular dynamics was done in the canonical (NVT) ensemble 
using the Verlet algorithm and a Nos\'e thermostat.
6501 steps with time-step
0.941~femtoseconds were taken for each simulation.  Each calculation
utilized 5 nodes on the University of Florida HiperGator cluster
(Gen.\ 3), with 64 cores per node (total of 320 cores) and 4~GB of
memory per core (1280 GB total).

\begin{figure}[ht]
  \begin{subfigure}[b]{0.49\textwidth}
  \centering
        \includegraphics[angle=0,scale=0.45]{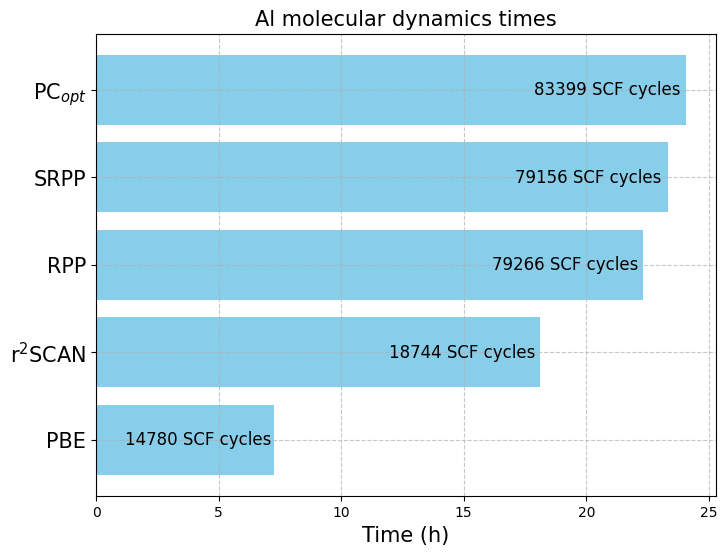}
        \caption{Bar chart showing total run times in hours; 
          total number of SCF cycles taken also noted.}
        \label{fig:DM_summary}
  \end{subfigure}

  \begin{subfigure}[b]{0.49\textwidth}
  \centering
       \includegraphics[angle=0,scale=0.45]{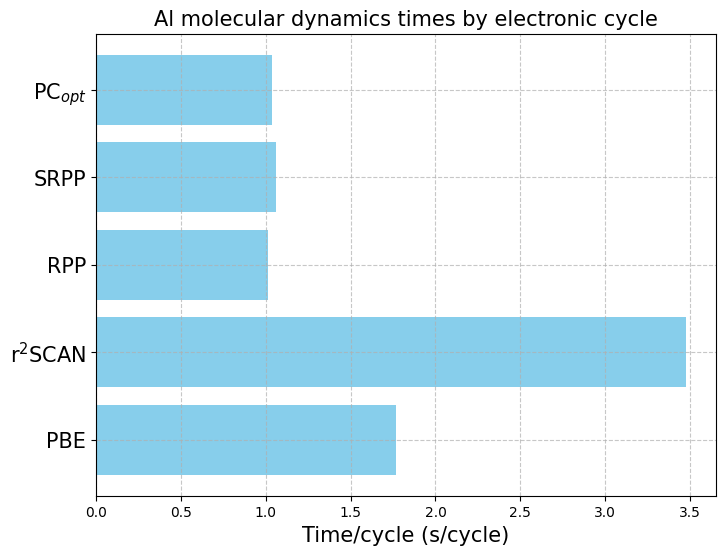}
       \caption{Time per electronic cycle, in seconds.}
       \label{fig:DM_TS}
   \end{subfigure}
   \caption{Timing statistics for 6,501 step AIMD simulation of liquid Al,
          using PBE, r$^2$SCAN, and r$^2$SCAN-L with several deorbitalization strategies.}
   \label{fig:DM}
\end{figure}

Fig.~\ref{fig:DM} gives the resulting timing statistics. Total times in hours
are in the upper panel, Fig.~\ref{fig:DM}(a), along with the
total number of SCF cycles. The time per SCF cycle
is shown in the lower panel Fig.~\ref{fig:DM}(b).
The outcome is that all of the deorbitalized functionals
are at a disadvantage with respect to the parent functional r$^2$SCAN
for total run time.  This is in contrast to what one would
have expected from the fcc Al equation of state results just discussed. 

Clearly, the problem is not in the time per SCF cycle.  As with the Al
equation of state, the best performers on a per-cycle basis are the
deorbitalized functionals, particularly RPP and SRPP. They out-perform
even PBE. But for the AIMD, the number of SCF cycles needed for those
deorbitalized functionals is almost prohibitively high, over four times the
number needed for the parent functional, and over five times that
needed for PBE.

This excessive SCF cycle count also is in marked contradiction to our
findings for the 55 solids. Recall Table~\ref{Table:changes_all}.  It
shows that for the lowest accuracy tolerances in cutoff energy and SCF
convergence (Case 6), the same as employed in the AIMD, the number of
cycles needed for the deorbitalized functionals and the parent
functional are roughly the same.  That clearly is not the case in the
AIMD driven by any of the deorbitalizers.

One way to characterize this problematic cycle count is that
convergence to self-consistency from the fully converged density of
the previous MD step is not that much quicker than convergence from an
LCAO density (12 cycles average per MD step for SRPP, in contrast to
14 cycles average over the test set and 18 to 21 for the equation of
state calculation for Al).  The behavior also is in contrast with the
dramatic reduction of PBE and r$^2$SCAN requirements to less than 3
cycles average per step.  This may be an unpleasant consequence of the
inclusion of the density Laplacian in the X functional and associated
sensitivity to seemingly small changes in the density. A peculiarity
is that a tolerance of $E_{\mathrm{diff}} =2.75 \times 10^{-5}$ needs
an average of 14 cycles to converge in the case of the 55 element test
set (see Table~\ref{Table:changes_all}) while, on the other hand, use
of $E_{\mathrm{diff}} = 1 \times 10^{-4}$ in the MD needs 18 cycles.
The difficulty is that though RMM-DIIS is extremely fast it is less
stable than the other solvers.

We have pursued two diagnostic follow-ups which indicate that the
issue is quite complex and not readily resolvable.  First, altering
the MD timestep should provide a quick consistency check regarding the
behavior.  We find remarkably that the
convergence issues, if anything, \textit{increase} as the timestep is
decreased.  See Table~\ref{table:MD_potim}. Since this obviously cannot be the case in the
limit of zero time-step, something peculiar is happening but we have
been unable to identify the cause.
\begin{table*}[h]
\centering
\caption{Total times for the molecular dynamics of Al using the SRPP deorbitalizer for different values of timestep.}
\renewcommand{\tabcolsep}{6pt}
\scalebox{0.9}{
\begin{tabular}{@{}lccccc@{}}
\toprule
XC functional                     & timestep (POTIM)   & Total time (s) &Number of SCF cycles & Time by cycle (s) & Average Temperature (K) \\ 
\midrule
\hline
\multirow{3}{*}{r$^2$SCAN-L(SRPP)}&  1.4               &  87897.82       & 78872   & 1.11              & 1023.021 \\
				  &  0.940793     	& 83935.23       & 79156   & 1.06              & 1022.869 \\
                                  &  0.5               & 95314.36       & 86477   & 1.10              & 1022.973 \\
\hline
 \multirow{2}{*}{r$^2$SCAN-L(RPP)}&  0.940793          & 80395.09       & 79266   & 1.01              & 1022.995 \\
                                  &  0.5               & 86879.67       & 87786   & 0.99              & 1023.052 \\
\hline
\multirow{2}{*}{r$^2$SCAN}       &  0.940793          & 65212.99       &  18744  & 3.48              & 1023.058 \\
                                  &  0.5               & 54600.70       & 13554   & 4.03              & 1023.056 \\
\hline
 \bottomrule
\end{tabular}}
\label{table:MD_potim}
\end{table*}
 
\begin{figure}[ht]
\centering
\caption{Number of SCF cycles to convergence per Molecular Dynamics step for different approximations, plotted for the first 100 MD time steps.} 
\includegraphics[angle=0,scale=0.33]{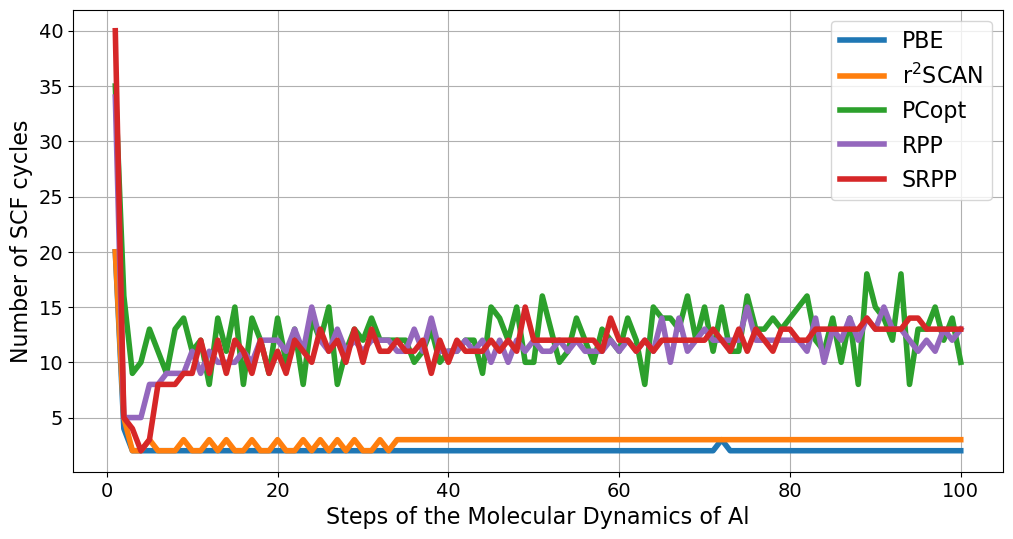}
\label{SCF_cycles_per_MD_step}
\end{figure}

Secondly, investigation of the number of SCF cycles needed as a
function of simulation time shows a rather intriguing trend. The RPP and
SRPP deorbitalized functionals start out with convergence
times that are quite competitive to r$^2$SCAN and PBE.  Only
gradually, over about 10 MD steps, do they settle down to the more
slowly converging behavior.  The slowdown in convergence is correlated
to increased fluctuations in convergence performance, 
symptomatic of an unresolved stability issue.
This behavior is documented in Fig.~\ref{SCF_cycles_per_MD_step}.  
Perhaps it is noteworthy that, of the three deorbitalizers,
the one with smoothest potential, SRPP has the smallest fluctuations in 
cycles-per-step. We have sought diagnostic insight from calculating 
 mean-square-displacements and
radial distribution functions for both T=1093K and 298K AIMD but without
much gain in understanding.  Plots of the results are in the Supplemental
Information. 

The behavior suggests that tendencies to instability in
the deorbitalized forms take some SCF cycles to show up and might be
ameliorated by taking an occasional step using, e.g. PBE, to find the
density.  We tried one such strategy. At each MD step, the
first five SCF cycles were done with PBE, then continued with
the selected deorbitalized functional. Unfortunately, this strategy
actually increased the number of SCF cycles instead of reducing them.

We also analyzed the Hellmann-Feynman forces during the course of this 
simulation. The detailed analysis is provided in Fig.~4 of the
Supplemental Information, which
shows the maximum, minimum, and median forces in the z direction for
the first 50 and 500 MD steps. No outliers are observed, and the median
values remain close to zero for all functionals tested. That
indicates a symmetric distribution of forces around zero, without
systematic bias introduced by the deorbitalized functionals.

\section{Discussion and Concluding remarks \label{sec:conclusion}}

Deorbitalizing a conventional meta-GGA XC functional is motivated by
two closely related goals.  First is 
to regain the interpretability provided by the locality of the KS potential.
Second is to achieve major computational speedups as a consequence of
that simplification without compromise of predictive accuracy.
Thus, the simultaneous focus of this paper is on issues of both functional
accuracy and speed. In that regard,  Laplacian-dependent deorbitalizers
are two-edged swords.  While providing useful  deorbitalization
accuracy, they can introduce highly spiky functions that can be impediments
to computational speed.

We have addressed those competing effects by introducing a smoothness
measure, $I^{deorb}$, for an XC potential to
motivate the use of  deorbitalizers (SRPP and SRPP2) designed with
smoothness in mind.  Both preserve
fundamental constraints respected by the recent OFR2 deorbitalization
of r$^2$SCAN and do not degrade norm satisfaction significantly.  The
result is improvement for both accuracy and timing objectives.  Our
timing study concentrates on approaches (gaussian basis sets,
plane-waves and PAW pseudopotentials) used in AIMD and high through-put
calculations, and should provide ballpark expectations for any such
code.  Tuning or adapting these codes for optimal performance for Laplacian-based
functionals, especially for AIMD, would be a welcome next step to
achieve the full promise of deorbitalization.

We have used the smoothness measure, $I^{deorb}$, to probe for and
control instability sources in the XC potential of deorbitalized
meta-GGAs.  
In consequence, SRPP and SRPP2 produce noticeably smoother kinetic and
exchange potentials than other deorbitalizers. Notably, for these  
smoothed variants of RPP,
the constraint compliance for slowly varying
densities imposed by the RPP
comes with a much smaller penalty in predictive performance
for highly inhomogeneous systems than one might expect from the
behavior of the RPP.  The smoothed functionals improve upon RPP
markedly for the G3 molecular test set, though the deorbitalized form
still is not competitive with r$^2$SCAN-L(PC\opt) for heats of
formation.  The (SRPP2) variant, though less effective than (SRPP) on
the molecular cases, is equal or better on the solids and delivers
better timing performance.  It is plausible that this gain is a
consequence of the r$^2$SCAN-L(SRPP2) X potential being smoother
than the SRPP-deorbitalized one. It also is plausible that smoothing along
  the lines we have presented would be helpful in reaching desired
  performance goals for deorbitalized finite-temperature functionals
  \cite{PhysRevB.105.L081109,feOFDFT2025}. 

The implications for timing performance are complicated and unfortunately
somewhat obscure.  For molecules, both the time per SCF cycle and the number
of cycles needed to reach convergence are essentially the same for
both the parent and deorbitalized functionals.
For fixed (static) solid
geometries, however, these particular deorbitalized
meta-GGAs significantly outperform the explicitly orbital dependent 
meta-GGA parent (used in gKS form) so far as time per SCF cycle is
concerned. While the deorbitalized forms (r$^2$SCAN-L(SRPP) or (SRPP2))
need more cycles to reach self-consistent convergence,
the increment is not so large as to offset the much shorter cycle
time.  As a result, these are much faster than r$^2$SCAN for
static lattice solid calculations, as seen in Tables
\ref{Table:soltime}, \ref{Table:changes_all} and~\ref{Table:altime}.

That advantage does not propagate into the one AIMD example we have
tried.  It appears that the density Laplacian dependence in the
deorbitalized functionals introduces instabilities that greatly
increase the number of SCF cycles needed in the AIMD context. It might
seem that the XC potential would be sensitive to small
perturbations in the density because of the Laplacian, so that potentials for
successive AIMD steps would be different enough to lead to
unusual variation in the density, and thus slower SCF convergence.
However, our limited testing with reduced AIMD step-time
is inconsistent with this argument.  Similarly, there is no 
obvious clue as to the cause of the problematic behavior in the average
SCF cycle time versus its spread.  Recall that Table~\ref{Table:soltime} shows
that r$^2$SCAN deorbitalized by SRPP, SRPP2, and RPP all have similar 
ratios of spread of
time per cycle to average time per cycle, and the ratio for SRPP
(2.35) is slightly below that for RPP (2.51).

Because Gaussian basis functions are themselves smooth and are
equivalent to fixed combinations of many plane waves, molecules
calculated with Gaussian basis sets behave, oddly enough,
like the cruder and faster plane-wave calculations, in the
sense that there is little distinction in timing among 
deorbitalizer forms.  Utilization of a smooth, analytical basis
apparently has a stabilizing effect that is a counterpart
to reduction of the energy cutoff in a plane-wave code.

Our limited AIMD calculations also display a third, somewhat
discomfiting  behavior.  As for the case of lower accuracy
calculations for the equation of state, deorbitalized functions
outperform the orbital-dependent parent on a per-cycle basis, while
smoothed and unsmoothed deorbitalizers have similar SCF convergence
rates.  But PBE and r$^2$SCAN perform the task specific to MD, namely
finding orbitals for a new set of nuclear positions starting from the
density for a modestly different set of positions, in two or three SCF
cycles.  In contrast, the deorbitalized forms used here take almost
the same number of cycles in starting from the previous converged
density for the previous AIMD step as they do when starting from, say, an LCAO
density.  An attempt to start at each step with a PBE density did
not help. The causes are undiagnosed. 

Finally, it may be advantageous to use the measure of 
noise we defined as an aid to construction of 
deorbitalizers with yet smoother potentials, either by optimizing the
switching function between iso-orbital and slowly-varying limits 
or strategically modifying ``appropriate norms" in the deorbitalizer or 
parent functional.  This is the focus of ongoing research.

\section*{Supplementary Material}
The following file \cite{SuppInfo} is available free of charge.
\begin{itemize}
  \item SRPP\_Timing.SuppInfo: In this file we provide a detailed, system-by-system tabulation of the numerical results of the test calculations against 
    standard molecular and solid test sets.  We also provide system-by-system
    SCF timing and SCF cycle count tabulations, corresponding to the  plot of SCF cycle
    count for each of the first 100 AIMD steps in Fig.~\ref{SCF_cycles_per_MD_step} above.  
   \item Plots of mean-squared displacements and radial distribution functions of Al from the various XC functionals driving AIMD at T=298K and 1023K.
\end{itemize}

\section*{Data Availability}
The data that support the findings reported in this paper are openly available at the Materials Cloud Archive.\cite{MCArchive}

\begin{acknowledgments}
  We thank Valentin V. Karasiev for assistance regarding AIMD calculations with VASP. Work supported by 
U.S. National Science Foundation grant DMR-1912618.
\end{acknowledgments}


\newpage
%

\end{document}